\documentclass{article}

\usepackage{arxiv}

\usepackage[utf8]{inputenc} 
\usepackage[T1]{fontenc}    
\usepackage{hyperref}       
\usepackage{url}            
\usepackage{booktabs}       
\usepackage{amsfonts}       
\usepackage{nicefrac}       
\usepackage{microtype}      
\usepackage{graphicx}       
\usepackage{natbib}         
\usepackage{doi}            
\usepackage{xcolor}         
\usepackage{CJK}            

\usepackage{indentfirst} 

\hypersetup{
  colorlinks=true,
  linkcolor=red,
  citecolor=orange,
  urlcolor=blue,
}

\graphicspath{{../figure}}

\title{Hairpin vortices and heat transfer in the wakes behind two hills \\with different scales
}

\author{\href{https://orcid.org/0000-0002-4875-8174}{\includegraphics[scale=0.06]{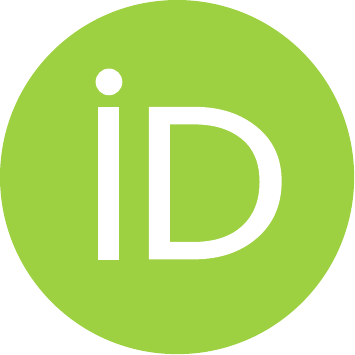}\hspace{1mm}Hideki Yanaoka (柳岡英樹)}
\thanks{Email address for correspondence: yanaoka@iwate-u.ac.jp} \\
	Department of Systems Innovation Engineering, \\
    Faculty of Science and Engineering, Iwate University, \\
    4-3-5 Ueda, Morioka, Iwate 020-8551, Japan \\
	\And
	Tatsuya Hamada (濵田 達哉) \\
    Mechanical Engineering, \\
    Graduate School of Engineering, Iwate University, \\
    4-3-5 Ueda, Morioka, Iwate 020-8551, Japan
	\texttt{} \\
}

\renewcommand{\shorttitle}{Hairpin vortices and heat transfer in the wakes behind two hills with different scales}

\makeatletter
\hypersetup{
pdfusetitle,
bookmarksnumbered,
pdftitle={\@title},
pdfsubject={},
pdfauthor={Hideki Yanaoka, Tatsuya Hamada},
pdfkeywords={Vortex dynamics, Hairpin vortex, Heat transfer, Unsteady flow, Boundary layer separation, Numerical simulation},
}
\makeatother

\begin{document}

\begin{CJK*}{UTF8}{ipxg} 
\maketitle
\end{CJK*}

\begin{abstract}
This study performed a numerical analysis of the hairpin vortex 
and heat transport generated by the interference of the wakes behind two hills 
in a laminar boundary layer. 
In the case of hills with the same scale, 
the interference between hairpin vortices in the wake is more intensive 
than in the different-scale hills. 
When the hills with different scales are installed, 
hairpin vortices with different scales are periodically shed. 
Regardless of the scale ratio of the hills, 
when the hill spacing in the spanwise direction is narrowed, 
the asymmetry of the hairpin vortex in the wake increases 
due to the interference between the wakes. 
At this time, the turbulence caused by the leg and the horn-shaped secondary vortex 
on the spanwise center side in the hairpin vortex increases, 
and heat transport around the hairpin vortex becomes active. 
In addition, the leg approaches the wall surface and removes high-temperature fluid 
near the wall surface over a wide area, 
resulting in a high heat transfer coefficient. 
These tendencies are most remarkable in the same-scale hills. 
In the case of hills with different scales, the heat transfer coefficient decreases 
because the leg on the spanwise center side in a small hairpin vortex 
does not develop downstream.
\end{abstract}


\section{Introduction}

In the flow around protuberance, 
separation bubbles occur ahead and behind the protuberance, 
and a coherent vortex structure exists in the wake. 
In such a flow field, heat, 
substance, and momentum are actively transported by the vortex. 
Therefore, many studies have investigated the flow and the wake 
around a single protuberance \citep{Acarlar&Smith_1987a,Sakamoto_et_al_1987,Yang_et_al_2001,Simpson_et_al_2002,Dong&Meng_2004,Yanaoka_et_al_2007a,Yanaoka_et_al_2007b,Yanaoka_et_al_2009} 
and multiple protuberances or cylinders \citep{Ko_et_al_1996,Sumner_et_al_1999,Meinders&Hanjalic_2002,Kurita&Yahagi_2008}.

When multiple protuberances are set in the flow, 
the wakes of the protuberances interfere with each other, 
resulting in complicated flows. 
Many studies have investigated interference between wakes behind multiple cylinders 
with the same or different diameters 
\citep{Ko_et_al_1996,Sumner_et_al_1999,Kurita&Yahagi_2008}. 
However, although the flow around the protuberance has been clarified 
in the previous study for multiple three-dimensional protuberances 
\citep{Meinders&Hanjalic_2002}, 
vortex structures in the wake and interference between wakes have not been investigated. 
Furthermore, in such a study on the flow around three-dimensional protuberances, 
protuberances with the same scale are used, 
and the wake behind protuberances with different scales has not been investigated.

In the wakes behind three-dimensional protuberances such as cubes 
\citep{Yanaoka_et_al_2007a}, 
trapezoidal tabs \citep{Dong&Meng_2004}, 
hemispheres \citep{Acarlar&Smith_1987a,Yang_et_al_2001}, 
and hills \citep{Simpson_et_al_2002,Yanaoka_et_al_2007b,Yanaoka_et_al_2009}, 
it is known that there are three-dimensional vortex structures such as hairpin vortices. 
So far, the authors \citep{Yanaoka_et_al_2007b} have performed 
a three-dimensional numerical analysis of the behavior and heat transfer 
of a hairpin vortex in the wake behind a single hill 
in a laminar boundary layer. 
It has been reported that hairpin vortices cause high turbulence in the wake 
and increase heat transfer. 
\citet{Li_et_al_2019} investigated the coherent structure and heat transfer 
in the turbulent boundary layer along a channel wall with a rib tabulator 
and found that multiple hairpin vortices occur, 
increasing the heat transfer coefficient and turbulence. 
When such three-dimensional protuberances are installed side by side in a flow field, 
the three-dimensional vortex structure in the wake is deformed 
by the interference between the wakes. 
Therefore, it is considered that the characteristics of turbulence 
and heat transfer change significantly compared to the case of a single protuberance. 
Furthermore, in the case of protuberances with different scales, 
the wakes with different scales interfere with each other. 
Thus, it is considered that the characteristics of the three-dimensional vortex structure, 
turbulence, and heat transfer in the wake are different 
from the characteristics of the flow around the protuberances with the same scale.

From the above viewpoints, 
this study investigates the interference between the wakes behind the hill-shaped protuberances 
in a laminar boundary layer 
to clarify the effects of the hill spacing and scale ratio 
on the hairpin vortex structure and heat transfer in the wakes behind two hills.

\section{Fundamental equation and numerical procedures}

\begin{figure}[!t]
\centering
\begin{minipage}{0.48\linewidth}
\centering
\includegraphics[trim=0mm 0mm 0mm 0mm, clip, width=70mm]{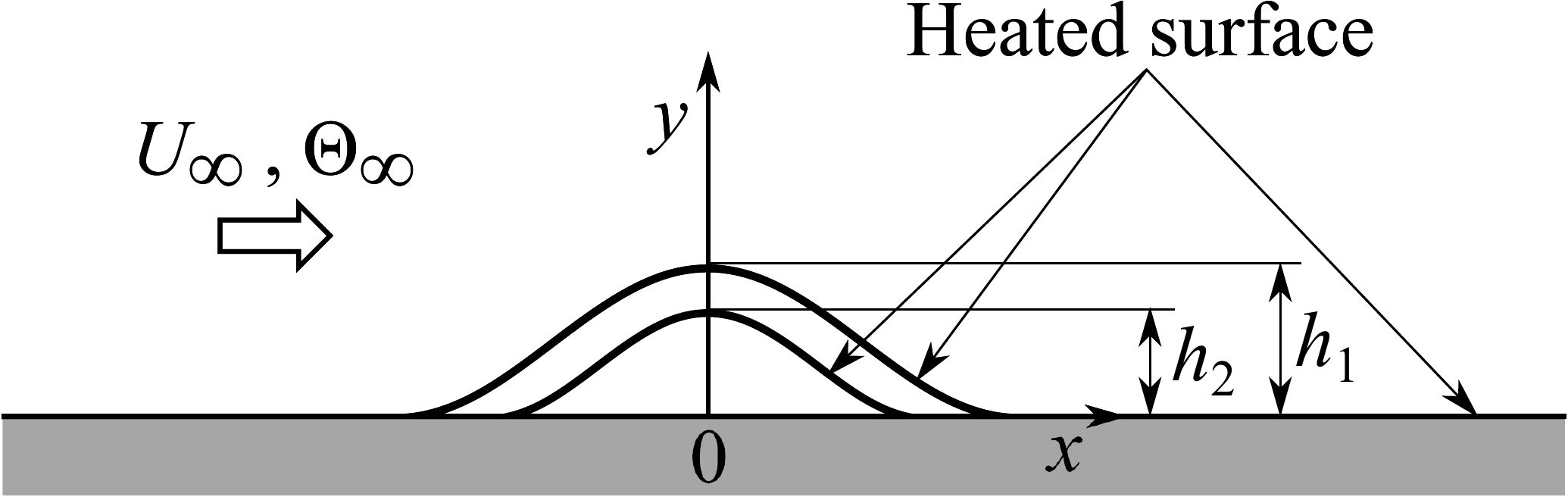} \\
(a) $x$-$y$ plane
\end{minipage}
\begin{minipage}{0.48\linewidth}
\centering
\vspace*{0.5\baselineskip}
\includegraphics[trim=0mm 0mm 0mm 0mm, clip, width=70mm]{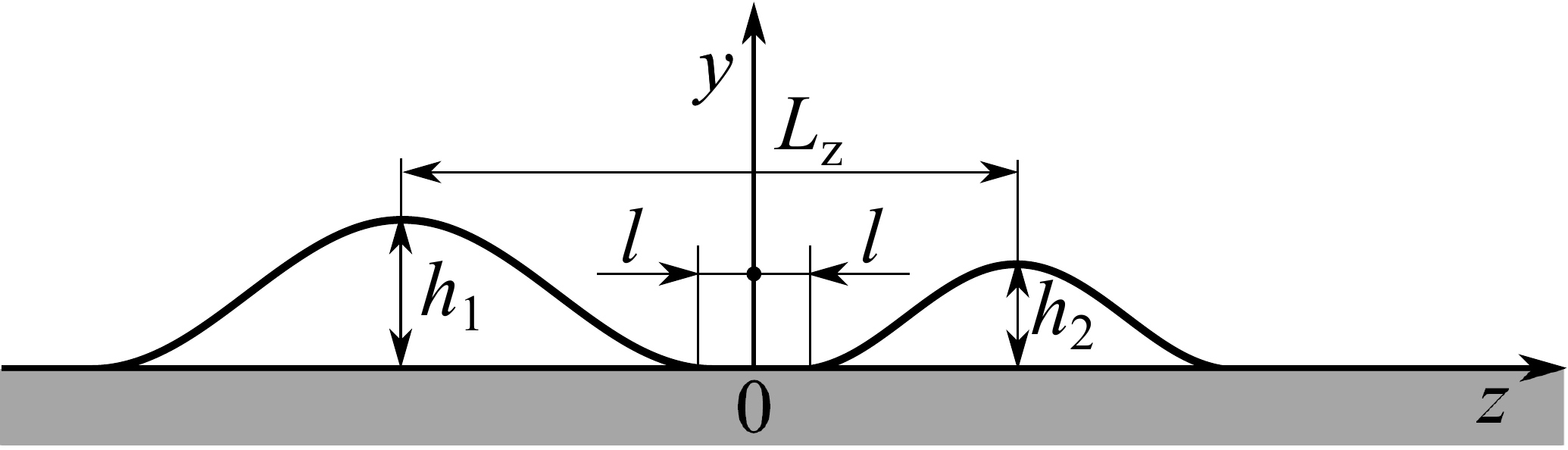} \\
(b) $y$-$z$ plane
\end{minipage}
\caption{Flow configuration and coordinate system.}
\label{coordinate}
\centering
\end{figure}

Figure \ref{coordinate} shows the flow configuration and coordinate system. 
The origin is on the wall surface, and the $x$-, $y$-, and $z$-axes are 
the streamwise, cross-streamwise, and spanwise directions, respectively. 
The velocities in these directions are denoted as $u$, $v$, and $w$, respectively, 
and the temperature is denoted as $\theta$. 
A uniform flow originates upstream and a laminar boundary layer develops downstream. 
The free stream velocity and temperature are denoted as $U_{\infty}$ and $\Theta_{\infty}$, respectively. 
In this study, similar to \citet{Acarlar&Smith_1987a}, 
a protuberance is placed in the laminar boundary layer, 
and a hairpin vortex is generated by using flow separation. 
The protuberance has the same shape as that used in the experiment of \citet{Simpson_et_al_2002}. 
The height shapes of the two hills are given as follows:
\begin{equation}
   \frac{y(r_k)}{h_1} = - \frac{1}{6.04844} \left[ J_0(\Lambda) I_0
   \left (\Lambda \frac{r_k}{a_{k}} \right) - I_0(\Lambda) J_0
   \left (\Lambda \frac{r_k}{a_{k}} \right) \right], 
\end{equation}
where $\Lambda=3.1926$, $r_k$ ($k=1$, 2) is the radius, 
$h_k$ is the hill height, $a_k=2h_k$ is the hill radius, 
$J_0$ is the Bessel function of the first kind, 
and $I_0$ is the modified Bessel function of the first kind. 
The vertices of two hills are placed on the $z$-axis, 
and the distance between two hills in the spanwise direction is $L_z$. 
The radius $r_k$ is given, respectively, as follows:
\begin{equation}
   r_1 = \sqrt{x^2 + (z + 2 h_{1} + l)^2} \quad \mbox{if} \quad (z<0),
\end{equation}
\begin{equation}
   r_2 = \sqrt{x^2 + (z - 2 h_{2} - l)^2} \quad \mbox{if} \quad  (z \geq 0),
\end{equation}
where $l$ is the distance from the bottom of the hill to the origin 
and is given as $l = (L_z - a_1 - a_2)/2$.

This study deals with the three-dimensional flow of an incompressible viscous fluid 
with constant physical properties. 
The continuity, Navier-Stokes, and energy equations in Cartesian coordinates 
are transformed into those to a general coordinate system. 
The governing equations are solved using the simplified marker and cell method 
\citep{Amsden&Harlow_1970} extended to the collocated grid. 
The pressure interpolation proposed by \citet{Rhie&Chow_1983} is used 
to remove spurious errors. 
The Crank--Nicolson method is applied to discretize the time derivative 
and then time marching is performed. 
The second-order central difference scheme is used to discretize the spatial derivative.

\section{Details of calculation}

In this study, we consider the cases where the scale ratio of two hills is $h_2/h_1=1$ and 7/8, 
and the hill spacing is $L_z=4h_1$ and $5h_1$. 
We adjusted the scale ratio so that a small-scale hairpin vortex was formed 
behind the hill at $z/h_1 \ge 0$ for $L_z=4h_1$ and set $h_2/h_1=7/8$. 
In the smaller scale ratio, the hairpin vortex was not generated. 
The calculation region is from $-10h_1$ to $25h_1$ in the $x$-direction 
and 0 to $8h_1$ in the $y$-direction. 
To unify the distance from the center of the hill to the boundary in the spanwise direction, 
for $L_z=4h_1$ and $5h_1$ of $h_2/h_1=1$, 
the calculation region in the $z$-direction is $-9h_1$ to $9h_1$ 
and $-9.5h_1$ to $9.5h_1$, respectively. 
For $L_z=4h_1$ and $5h_1$ of $h_2/h_1=7/8$, 
the region in the $z$-direction is $-9.125h_1$ to $9.125h_1$ 
and $-9.625h_1$ to $9.625h_1$, respectively.

For the boundary conditions of the velocity field, 
the Blasius velocity profile is given at the inlet, 
and convective boundary conditions are used at the outlet. 
No-slip boundary conditions are applied on the walls. 
Slip boundary conditions are assumed at the upper boundary away from the bottom surface 
and at the boundary in the spanwise direction. 
For the boundary condition of the temperature field, 
a uniform temperature is given at the inlet, 
and the convective boundary condition is used at the outlet. 
The hill surface and bottom wall are heated with a constant heat flux. 
A zero gradient of the temperature is assumed at the upper boundary 
and in the spanwise direction.

The numerical calculations are performed under the Reynolds number $Re=500$ 
defined by $h_1$ and $U_{\infty}$, and the Prandtl number $Pr=0.7$. 
The velocity boundary layer thickness $\delta$ at the inlet is set 
so that the thickness $\delta$ becomes $\delta_0/h_1=1$ at $x/h_1=0$ without the hill. 
These conditions are the same as the previous calculation \citep{Yanaoka_et_al_2007b}.

To confirm the grid dependency on the calculation result, 
we use the four grids for $h_2/h_1=1$ and three grids for $h_2/h_1=7/8$. 
For $h_2/h_1=1$, grid points of $181 \times 56 \times 133$ (grid1), 
$256 \times 70 \times 151$ (grid2), $300 \times 80 \times 169$ (grid3), 
and $335 \times 90 \times 179$ (grid4) are used. 
For $h_2/h_1=7/8$, grid points of $181 \times 56 \times 133$ (grid1), 
$256 \times 70 \times 151$ (grid2), and $335 \times 90 \times 179$ (grid3) are used. 
The grids are dense near the lower wall surface and hills. 
For each $h_2/h_1$, we investigated the influence of the number of grid points 
on the calculation results for $L_z=4h_1$, 
where the turbulence is the highest. 
We will describe the grid dependency later. 
In this study, to clarify the turbulence due to the vortex structure in more detail, 
the calculation results using grid4 and grid3 are mainly shown for $h_2/h_1=1$ and 7/8, respectively.

\section{Results and discussion}

\subsection{Flow around hills}

\begin{figure}[!t]
\centering
\includegraphics[trim=0mm 0mm 0mm 0mm, clip, width=70mm]{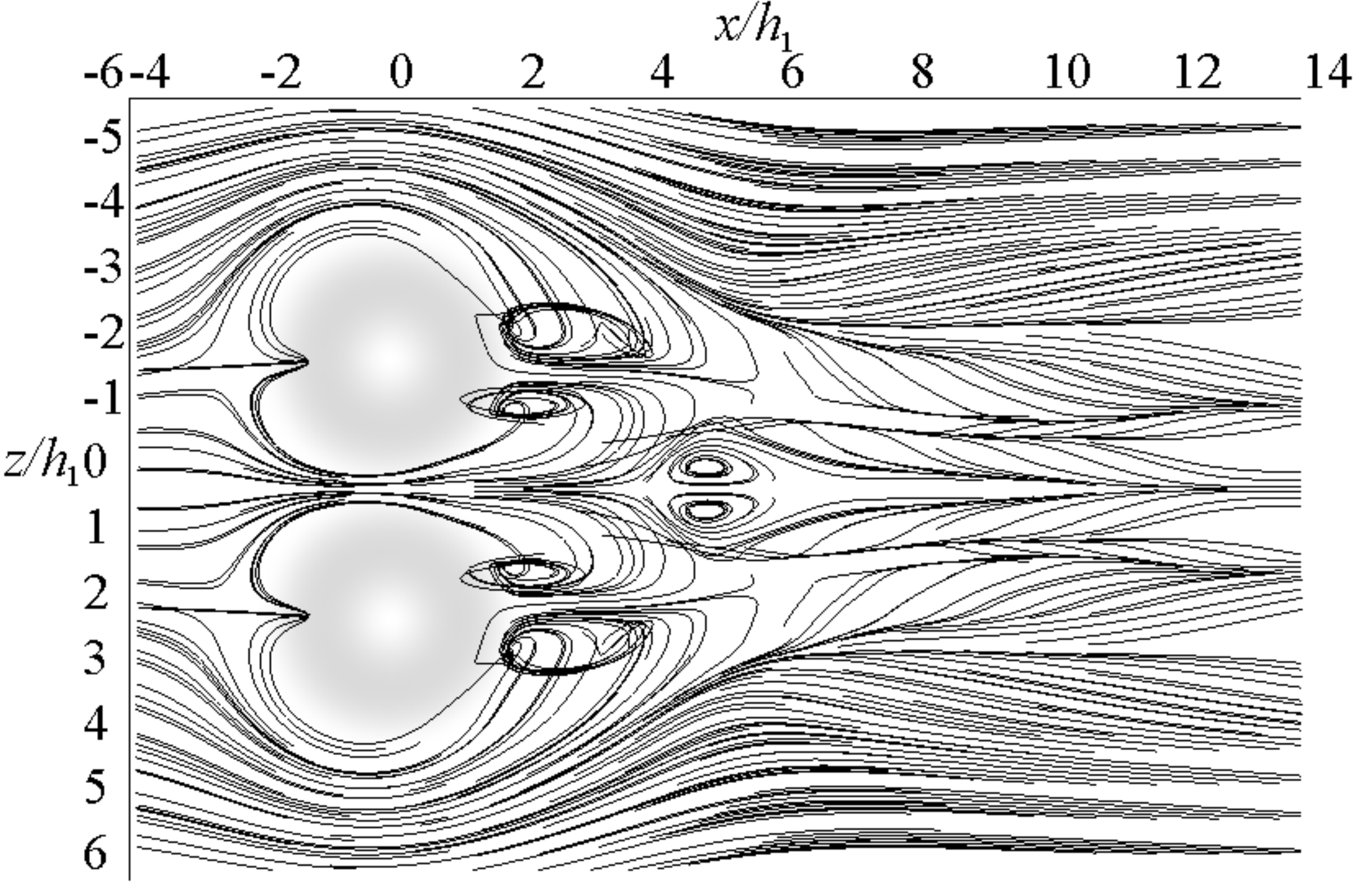}
\caption{Time-averaged streamlines at $y/h_1=0.1$ for $h_2/h_1=1$ and $L_z=4h_1$.}
\label{stream_top}
\end{figure}

Figure \ref{stream_top} shows the streamlines of the time-averaged flow 
in the $x$-$z$ cross-section for $h_2/h_1=1$ and $L_z=4h_1$. 
The results of other conditions were similar to the results of $h_2/h_1=1$ and $L_z=4h_1$, 
so they are omitted here. 
A pair of vortices are formed behind the hill. 
This vortex pair is similar to the CVP (Counter-rotating Vortex Pair) 
observed in the visualization experiment on the flow around a tab-shaped obstacle 
by \citet{Elavarasan&Meng_2000}.

As a contraction flow occurs between the hills and the pressure decreases, 
the fluid is drawn toward the center $z/h_1=0$ of the spanwise direction, 
and the separation bubbles move to the spanwise center. 
The movement of the separation bubble to the spanwise center reduces 
the size of the recirculation region on the spanwise center side 
compared to the result for a single hairpin vortex \citep{Yanaoka_et_al_2007b}. 
As a result, the vortex pair behind the hill becomes asymmetric 
to the center plane of the hill.

\begin{figure}[!t]
\centering
\vspace*{-0.5\baselineskip}
\begin{minipage}{0.48\linewidth}
\centering
\includegraphics[trim=0mm 0mm 0mm 0mm, clip, width=75mm]{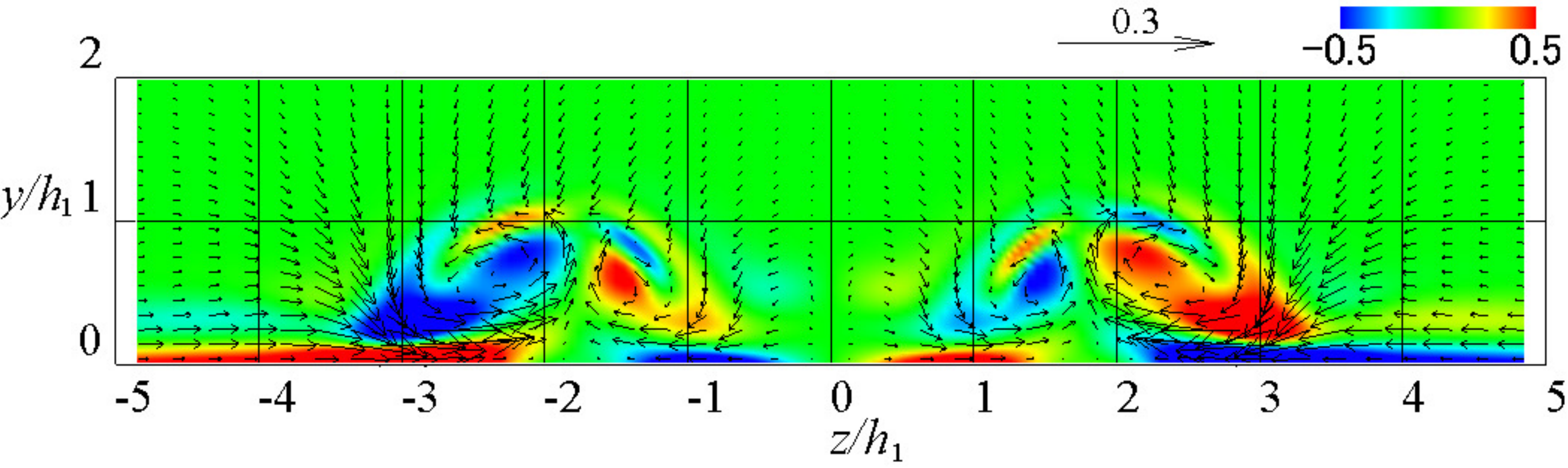}\\
(a) $h_2/h_1=1$ and $L_z=4h_1$
\end{minipage}
\begin{minipage}{0.48\linewidth}
\centering
\includegraphics[trim=0mm 0mm 0mm 0mm, clip, width=75mm]{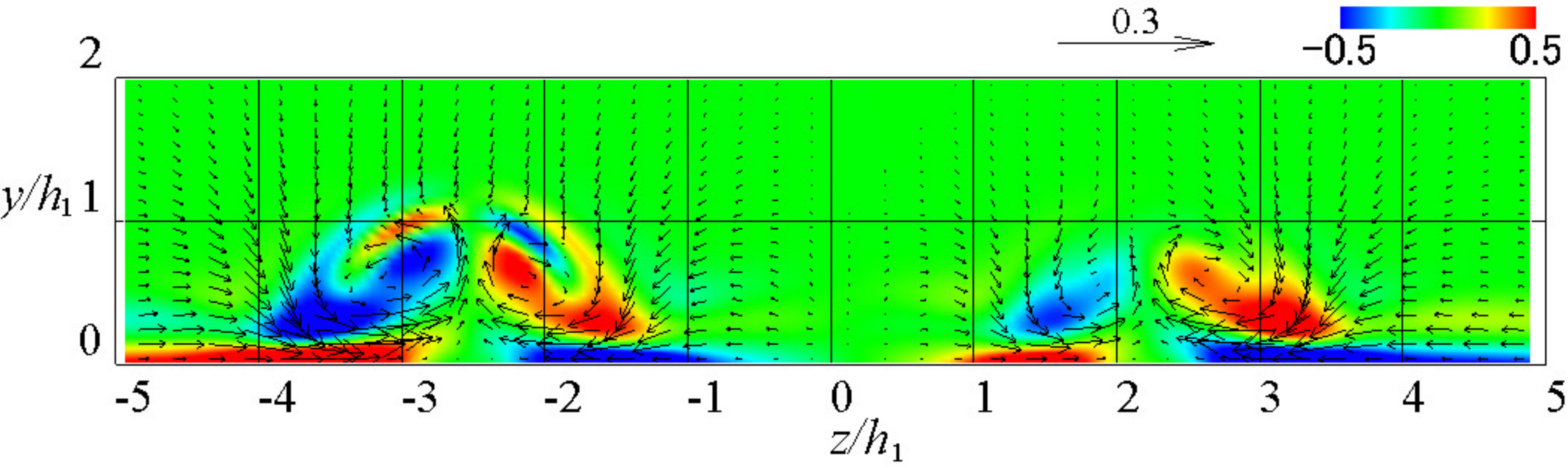}\\
(b) $h_2/h_1=7/8$ and $L_z=5h_1$
\end{minipage}
\caption{
Streamwise vorticity contours and velocity vectors in $y$-$z$ plane at $x/h_1=3.0$.}
\label{vor_vec}
\end{figure}

Figure \ref{vor_vec} shows the instantaneous streamwise vorticity contours 
and velocity vectors in the $y$-z plane. 
Under each condition, vortex pairs that rotate in opposite directions 
are asymmetrically formed behind the hills. 
In all cases, vorticity is low on the spanwise center side of the vortex pair, 
and flow circulation weakens. 
For each $h_2/h_1$, when the hill spacing is narrowed, 
the magnitude of the vorticity on the spanwise center side in the vortex pair 
at $z/h_1 \ge 0$ becomes lower. 
This tendency is remarkable for $h_2/h_1=7/8$ and $L_z=4h_1$.

\subsection{Vortex structures in the wake}

\begin{figure}[!t]
\centering
\vspace*{-0.4\baselineskip}
\begin{minipage}{0.48\linewidth}
\centering
\includegraphics[trim=0mm 0mm 0mm 0mm, clip, width=75mm]{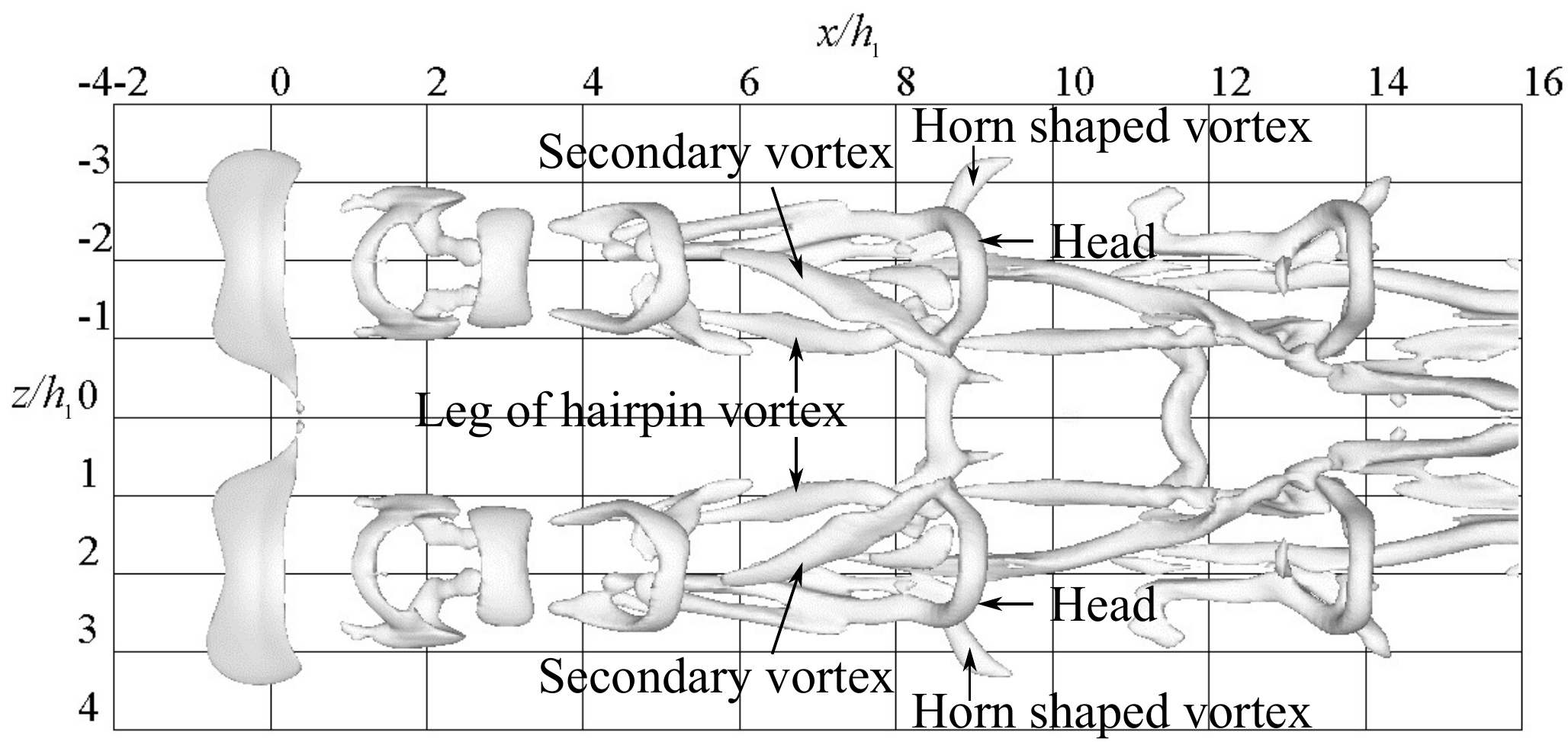} \\
(a) $h_2/h_1=1$, $L_z=4h_1$
\end{minipage}
\begin{minipage}{0.48\linewidth}
\centering
\includegraphics[trim=0mm 0mm 0mm 0mm, clip, width=75mm]{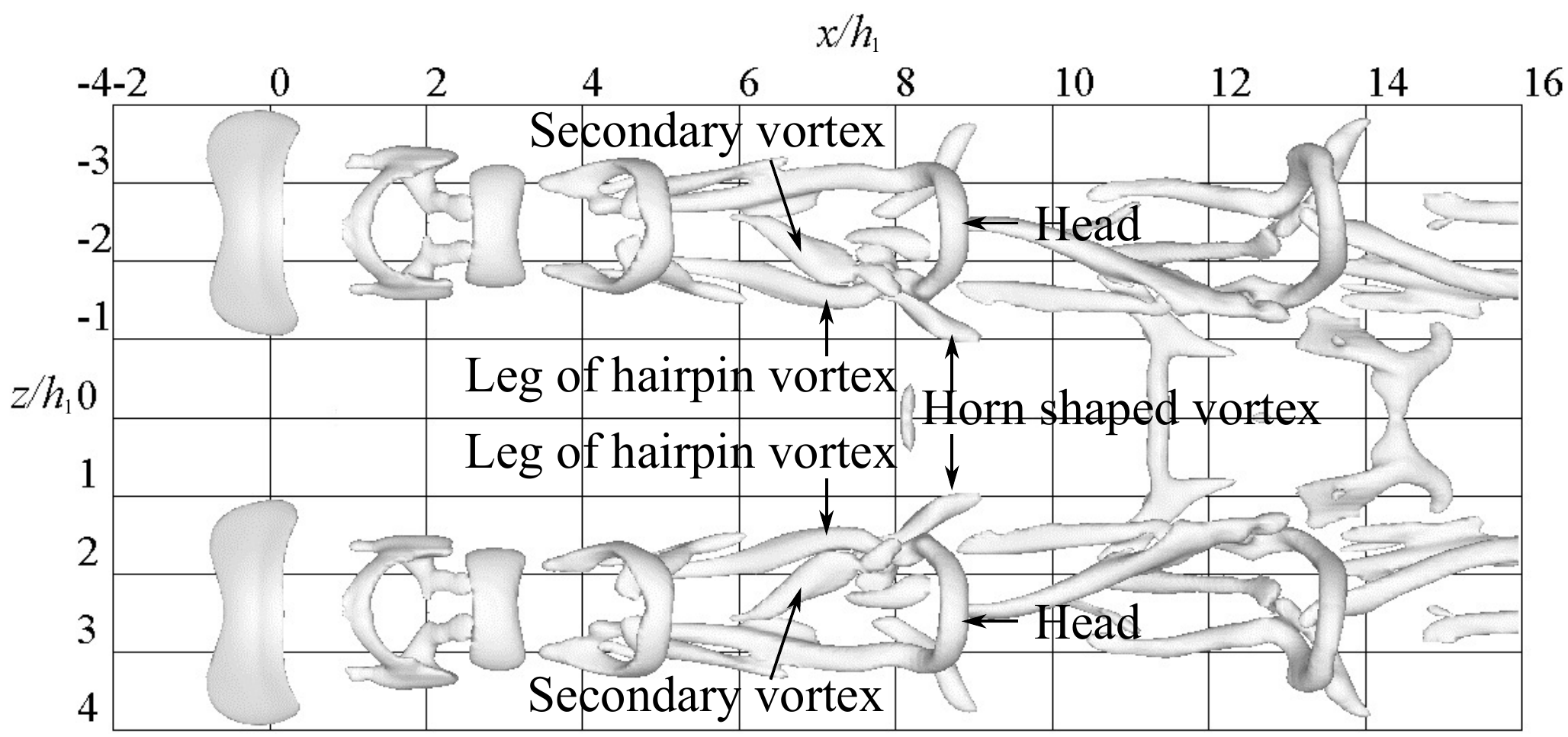} \\
(b) $h_2/h_1=1$, $L_z=5h_1$
\end{minipage}
\begin{minipage}{0.48\linewidth}
\centering
\vspace*{0.5\baselineskip}
\includegraphics[trim=0mm 0mm 0mm 0mm, clip, width=75mm]{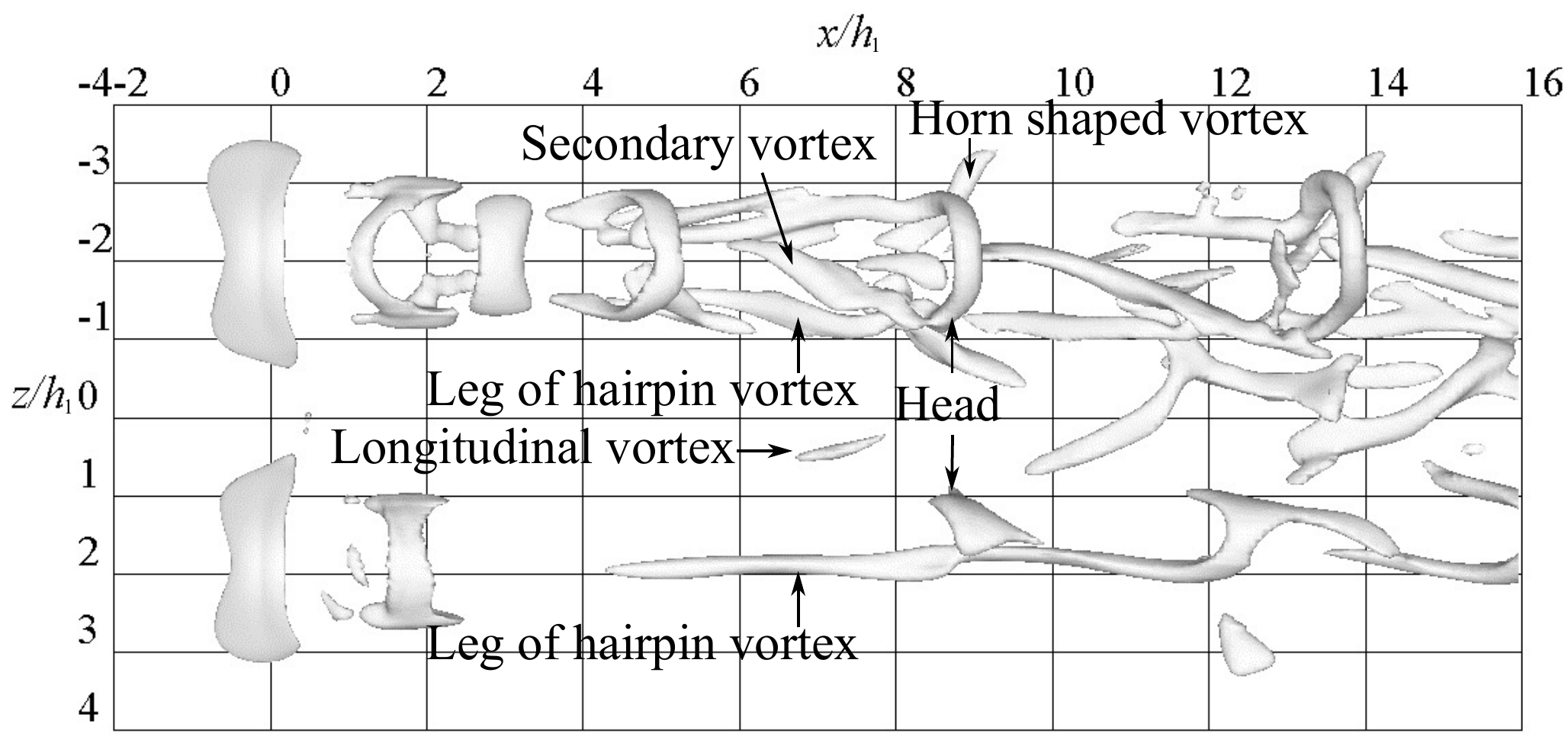} \\
(c) $h_2/h_1=7/8$, $L_z=4h_1$
\end{minipage}
\begin{minipage}{0.48\linewidth}
\centering
\vspace*{0.5\baselineskip}
\includegraphics[trim=0mm 0mm 0mm 0mm, clip, width=75mm]{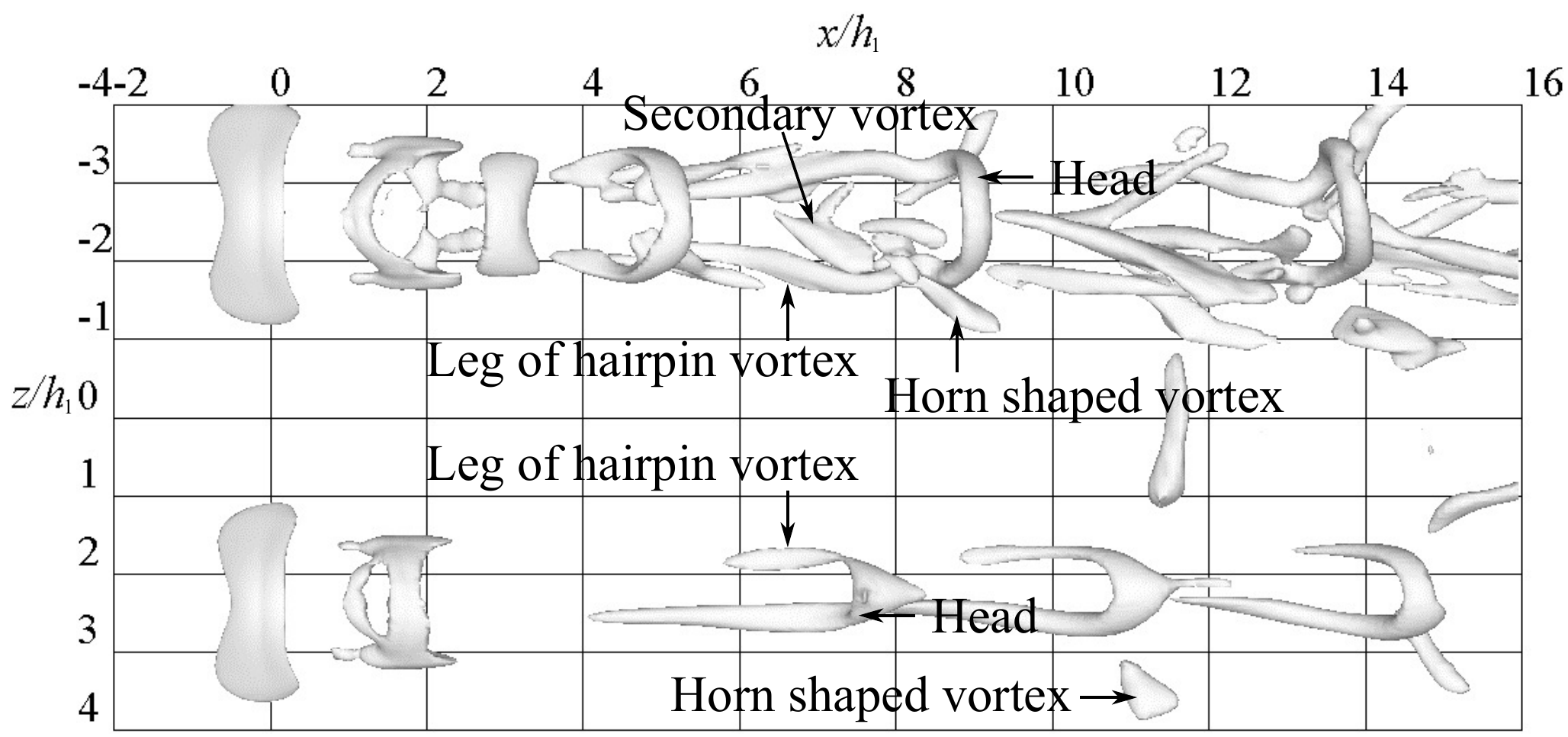} \\
(d) $h_2/h_1=7/8$, $L_z=5h_1$
\end{minipage}
\caption{Isosurface of curvature of equipressure surface: Isosurface value is $-7/h_1$.}
\label{curv_top}
\end{figure}

Figure \ref{curv_top} shows the isosurface of the curvature calculated 
from the equipressure surface to clarify the vortex structure 
existing in the flow field. 
To compare the instantaneous vortex structures in the same state for each condition, 
we show the flow field when the head center of a hairpin vortex shed from the hill 
at $z/h_1<0$ passes through $x/h_1=5$. 
The position where the pressure of the head became minimum was defined as the head center. 
In the flow field, there exist the head and leg of a hairpin vortex, 
the secondary vortex, the horn-shaped secondary vortex, 
and the longitudinal vortex. 
In each figure, these vortex structures are labeled. 
Under all conditions, the shear layer separated at the top of the hill 
becomes unstable downstream and rolls up into a vortex. 
This vortex grows into a hairpin vortex and is periodically shed downstream.

For $h_2/h_1=1$, the hairpin vortices around $x/h_1=4-5$ are similar to 
the existing result \citep{Yanaoka_et_al_2007b}. 
However, the hairpin vortices existing at $x/h_1=6-9$ become asymmetrical 
due to the interference of the vortices with each other, 
which is different from the previous result \citep{Yanaoka_et_al_2007b}. 
Around $x/h_1=6-8$, a secondary vortex is generated between both legs of the hairpin vortex. 
This secondary vortex develops above the leg on the spanwise center side 
as it moves downstream. 
A horn-shaped secondary vortex occurs under the hairpin vortex. 
These horn-shaped secondary vortices have also been observed 
in the previous study \citep{Yanaoka_et_al_2007b}. 
The horn-shaped secondary vortices on the spanwise center side spread 
in the spanwise direction with time and are coupled to each other downstream. 
In the case of $L_z=4h_1$, 
it can be seen that horn-shaped secondary vortices are connected around $x/h_1=8.5$. 
As the hill spacing at $L_z=5h_1$ is wider than at $L_z=4h_1$, 
it is observed that they are combined at $x/h_1=11.5$ further downstream. 
In existing studies using two-dimensional cylinders \citep{Sumner_et_al_1999,Kurita&Yahagi_2008}, 
it was observed that the flow field was symmetrical to the spanwise direction 
when the cylinder spacing was wide. 
On the other hand, it was observed that the flow field became asymmetric 
when the cylinder spacing was narrow. 
A similar tendency appears in this study.

For $h_2/h_1=7/8$, the hairpin vortex shed from the hill at $z/h_1\ge0$ is smaller than 
the hairpin vortex at $z/h_1<0$. 
The small-scale hairpin vortex interferes with the large-scale hairpin vortex, 
so they have a stronger asymmetry than the hairpin vortex for $h_2/h_1=1$. 
Furthermore, when the hill spacing is narrowed, 
the leg on the spanwise center side of the small hairpin vortex does not extend. 
As shown in Fig. \ref{vor_vec}(d), behind the hill at $z/h_1\ge0$, 
the vorticity magnitude on the spanwise center side decreases 
due to the interference between the wakes, 
and the rollup of the vortex is weakened. 
Therefore, the leg on the spanwise center side in the small hairpin vortex 
does not develop downstream and decays immediately. 
As a result, that leg does not extend. 
For $h_2/h_1=7/8$ and $L_z=5h_1$, unlike other conditions, 
the secondary vortex generated between both legs of the hairpin vortex at $z/h_1<0$ spreads 
above both legs in the spanwise direction when it moves downstream. 
For $h_2/h_1=7/8$ and $L_z=4h_1$, a longitudinal vortex can be seen near the spanwise center. 
This longitudinal vortex is secondarily generated by large- and small-scale hairpin vortices 
and extends downstream toward the large-scale hairpin vortex. 
Therefore, it is considered that interference occurs downstream, 
even between this longitudinal vortex and the large-scale hairpin vortex.

The vortex shedding frequency of such a hairpin vortex was $f=0.158U_{\infty}/h_1$ 
under all conditions. 
This value is close to the results, $f=0.159U_{\infty}/h_1$ and $f=0.163U_{\infty}/h_1$, 
at $Re=500$ by the authors \citep{Yanaoka_et_al_2007a,Yanaoka_et_al_2007b}. 
In addition, this result exists in the range of the experimental results, 
$f=0.13-0.25U_{\infty}/h_1$ and $f=0.12-0.17U_{\infty}/h_1$, at $Re=500$ 
by \citet{Acarlar&Smith_1987a} and \citet{Sakamoto_et_al_1987}, respectively.

\begin{figure}[!t]
\centering
\begin{minipage}{0.48\linewidth}
\centering
\includegraphics[trim=0mm 6mm 0mm 5mm, clip, width=75mm]{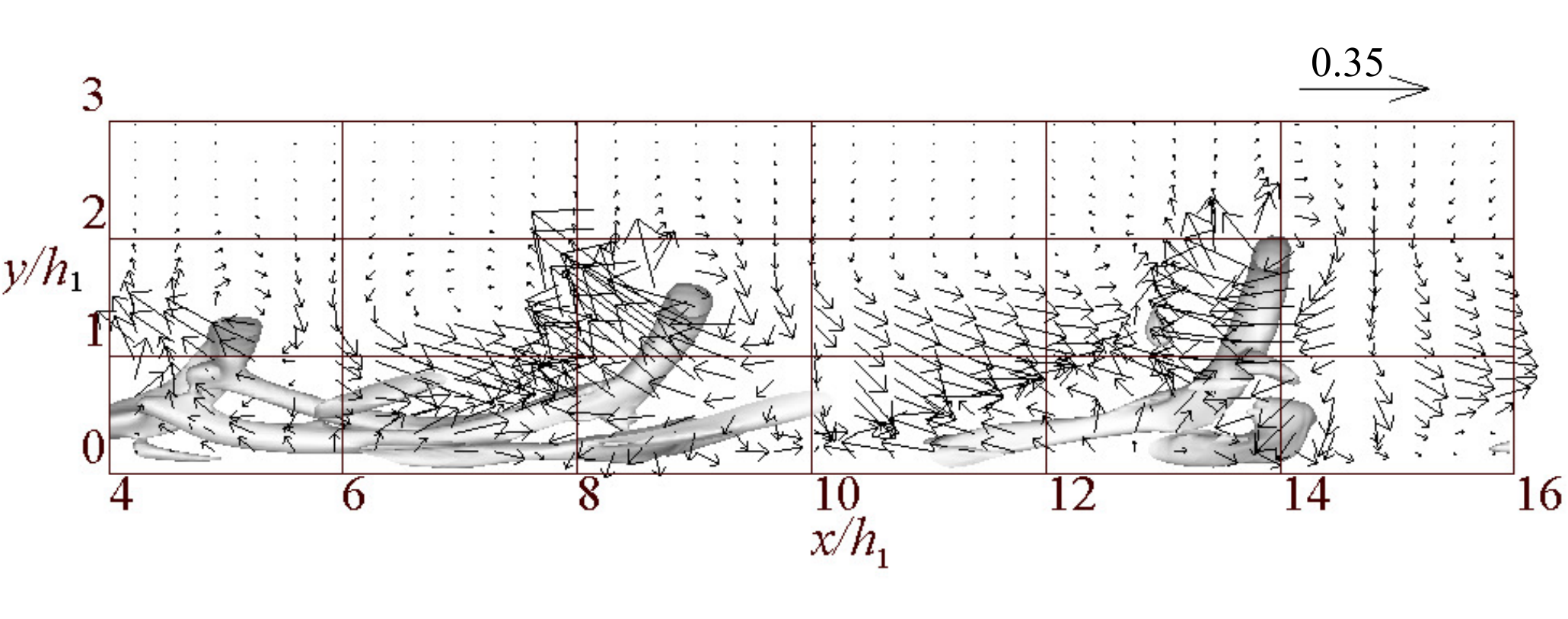} \\
(a) $h_2/h_1=1$, $L_z=4h_1$ ($z/h_1=-2.0$)
\end{minipage}
\begin{minipage}{0.48\linewidth}
\centering
\vspace*{0.05\baselineskip}
\includegraphics[trim=0mm 6mm 0mm 5mm, clip, width=75mm]{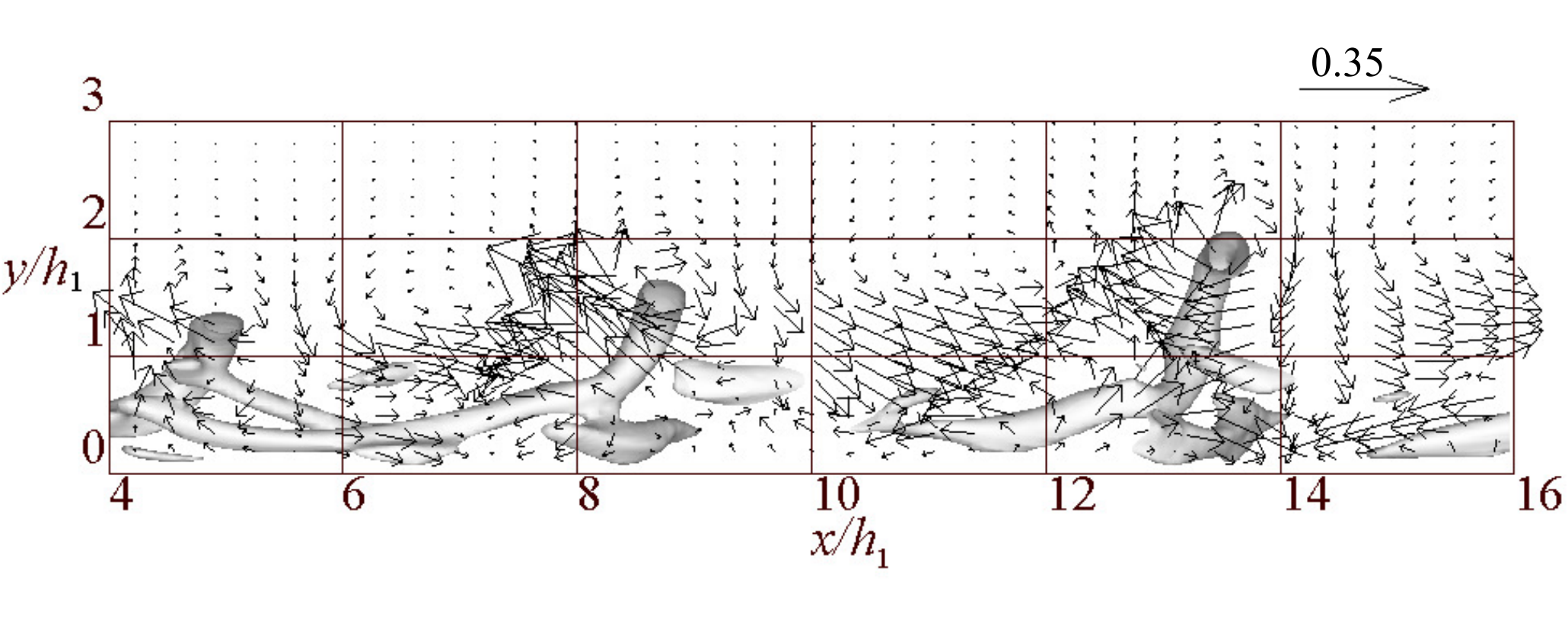} \\
(b) $h_2/h_1=1$, $L_z=5h_1$ ($z/h_1=-2.5$)
\end{minipage}
\begin{minipage}{0.48\linewidth}
\centering
\vspace*{0.5\baselineskip}
\includegraphics[trim=0mm 6mm 0mm 5mm, clip, width=75mm]{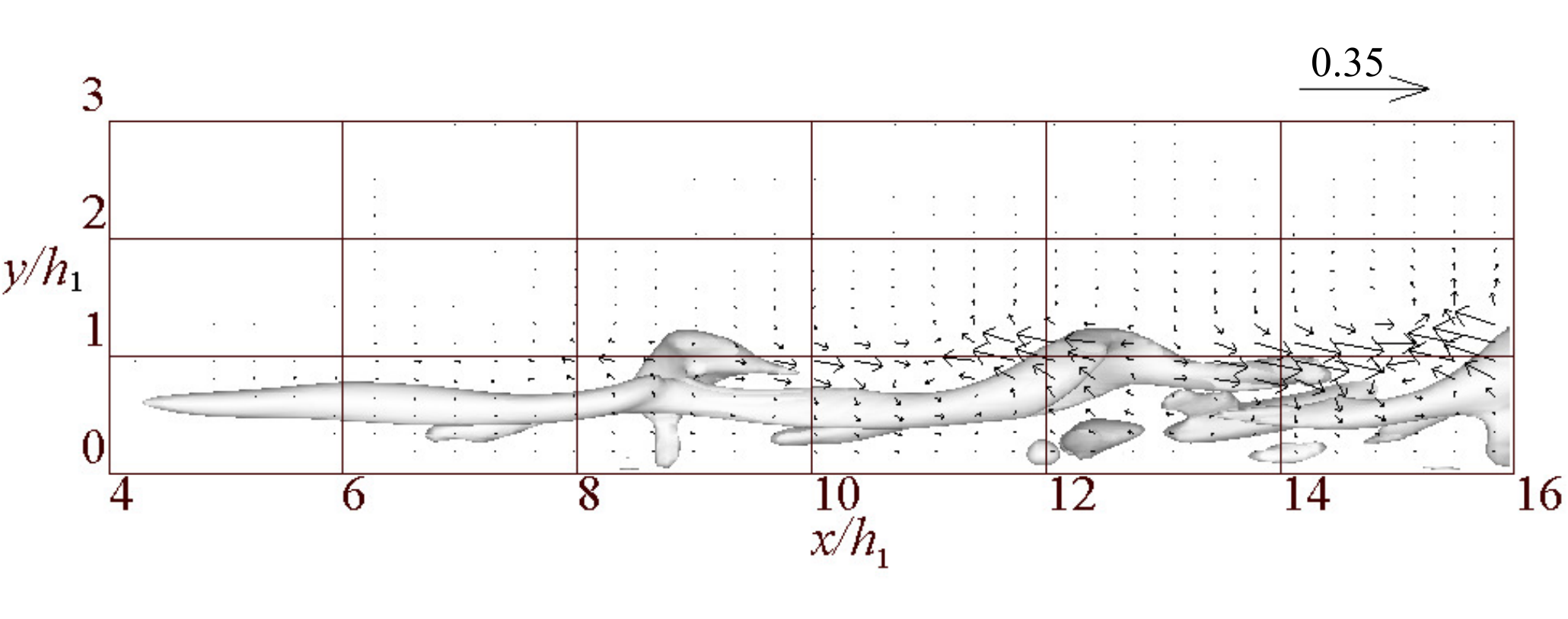} \\
(c) $h_2/h_1=7/8$, $L_z=4h_1$ ($z/h_1=1.4$)
\end{minipage}
\begin{minipage}{0.48\linewidth}
\centering
\vspace*{0.55\baselineskip}
\includegraphics[trim=0mm 6mm 0mm 5mm, clip, width=75mm]{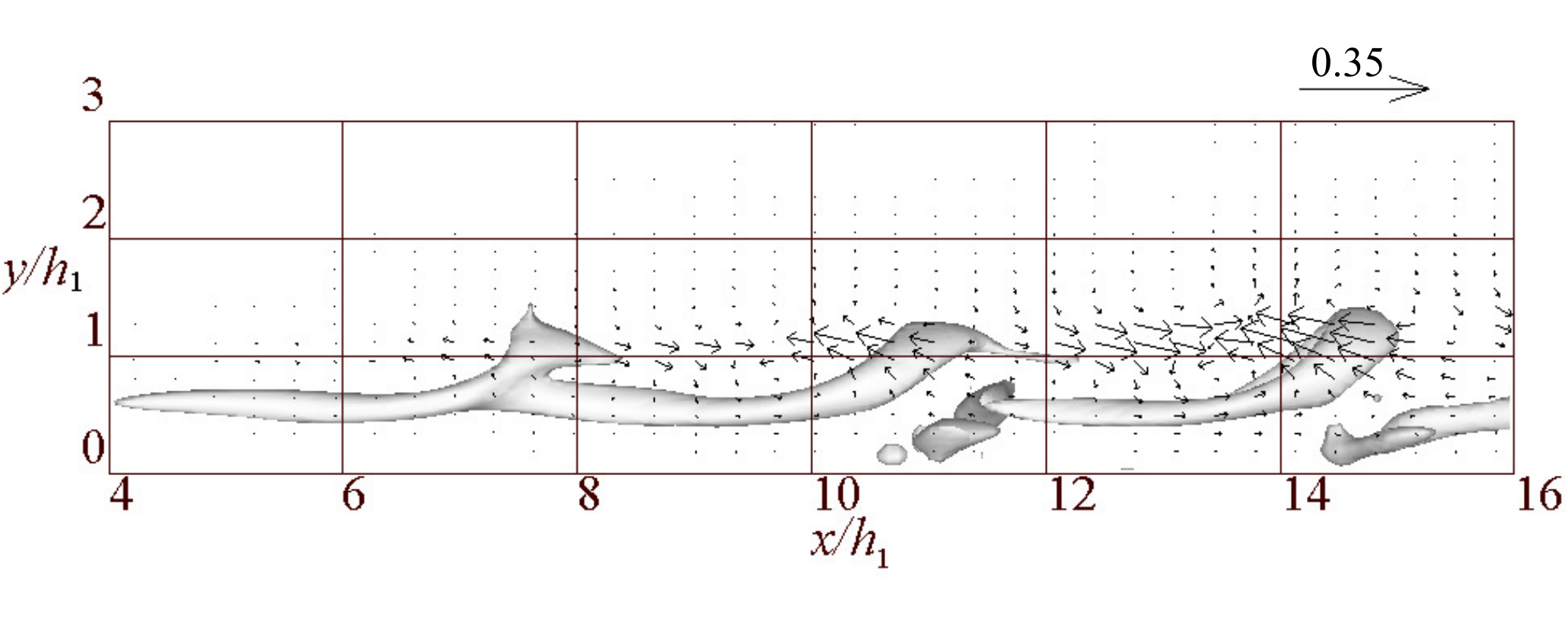} \\
(d) $h_2/h_1=7/8$, $L_z=5h_1$ ($z/h_1=2.2$)
\end{minipage}
\caption{Isosurface of curvature of equipressure surface and velocity fluctuation vectors in $x$-$y$ plane: 
Isosurface value is $-7/h_1$.}
\label{curv_vec}
\end{figure}

To observe the variations of the flow around the hairpin vortex with $h_2/h_1$ and $L_z$, 
Fig. \ref{curv_vec} shows the isosurface of curvature of the equipressure surface 
and the velocity fluctuation vectors in the $x$-$y$ plane. 
The results of $L_z=4h_1$ and $5h_1$ for $h_2/h_1=1$ show 
the central cross-sections of the hills at $z/h_1=-2.0$ and $-2.5$, respectively. 
The results of $L_z=4h_1$ and $5h_1$ for $h_2/h_1=7/8$ show 
the vicinity of the central cross-section of the hills at $z/h_1=1.4$ and 2.2, respectively. 
As the tendency of the wake at $z/h_1<0$ of $L_z=4h_1$ and $5h_1$ for $h_2/h_1=7/8$ 
is similar to the result of $L_z=4h_1$ and $5h_1$ for $h_2/h_1=1$, 
the result is omitted here. 
In each case, a clockwise flow is induced around the head of the hairpin vortex. 
The phenomenon of flow from the lower side of the hairpin vortex head 
to the upstream mainstream side is called Q2 ejection, 
and the flow toward the downstream wall surface in front of the hairpin vortex head 
is called Q4 sweep \citep{Yang_et_al_2001}.

For $h_2/h_1=1$, as shown around $y/h_1=1$ at $x/h_1=7-8$, 
an intensive shear layer occurs in the region 
where the flows generated by the Q2 ejection and Q4 sweep collide. 
This shear layer creates a secondary vortex between both legs. 
Around $x/h_1=14$ for $h_2/h_1=1$, 
a downward flow is generated by the horn-shaped secondary vortex. 
For $h_2/h_1=1$ and $L_z=5h_1$, 
an ascending flow is locally generated near the wall surface at $x/h_1=12$. 
This is due to the rotation of the legs of the hairpin vortex. 
When the hill spacing is narrowed, 
the asymmetry of the hairpin vortices increases due to the strong interference 
between the hairpin vortices. 
As a result, because the leg of the hairpin vortex is closer to the central cross-section of the hill, 
the rotation of the leg affects the flow field over a wide range. 
Therefore, for $h_2/h_1=1$ and $L_z=4h_1$, 
an ascending flow occurs over a wide area near the wall surface at $x/h_1=10-13$.

For $h_2/h_1=7/8$, 
as the velocity fluctuation around the head is smaller than 
the result for $h_2/h_1=1$, 
an intensive shear layer does not occur around the area 
where the flows generated by the Q2 ejection and Q4 sweep collide. 
Therefore, no secondary vortices are formed between both legs. 
Furthermore, the velocity fluctuation of the flow around the leg 
is smaller than that of the large-scale hairpin vortex generated 
under each condition.

\begin{figure}[!t]
\centering
\hspace*{0.25\baselineskip}
\begin{minipage}{0.48\linewidth}
\centering
\includegraphics[trim=19mm 0mm 25mm 0mm, clip, width=75mm]{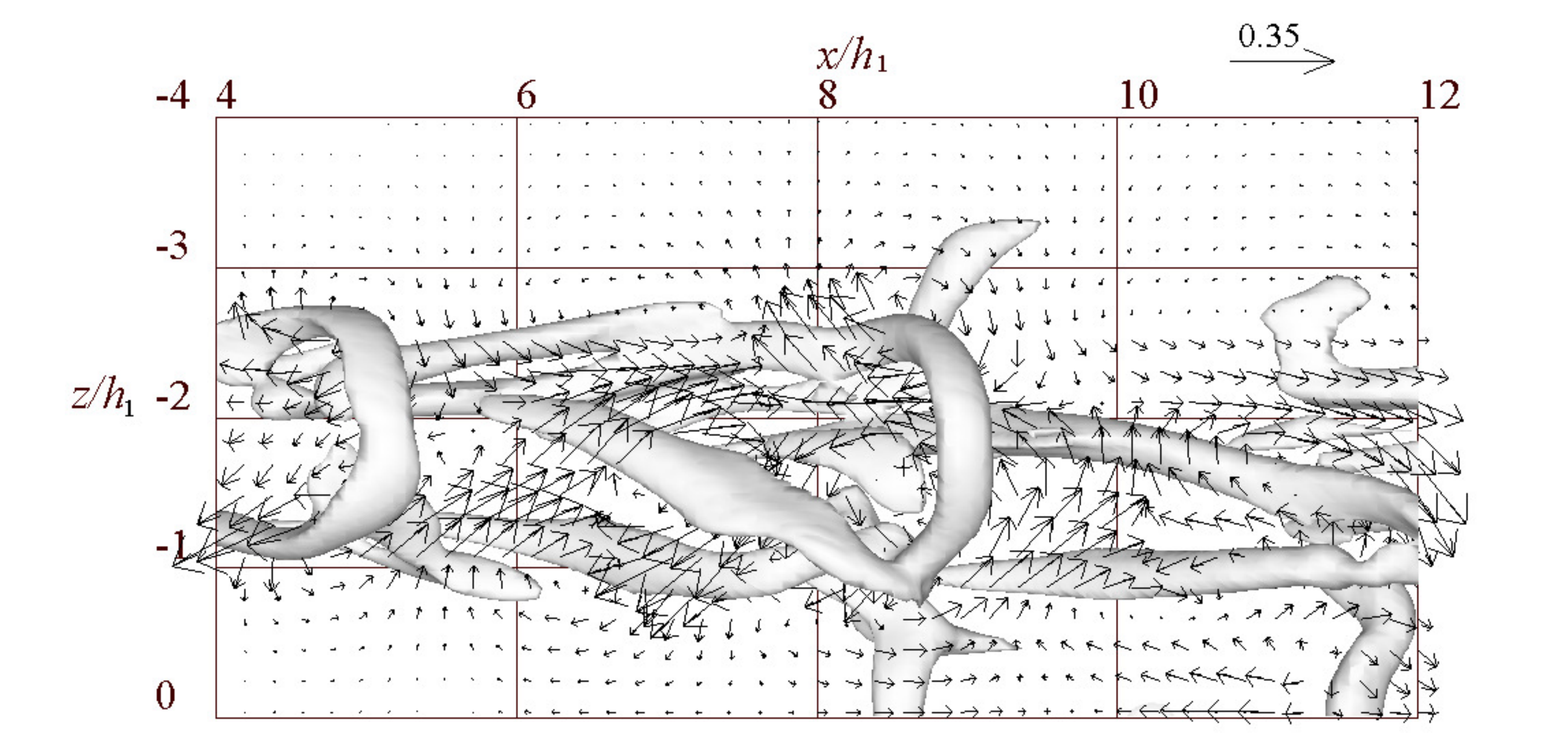} \\
(a) $h_2/h_1=1$, $L_z=4h_1$
\end{minipage}
\begin{minipage}{0.48\linewidth}
\centering
\includegraphics[trim=19mm 0mm 25mm 0mm, clip, width=75mm]{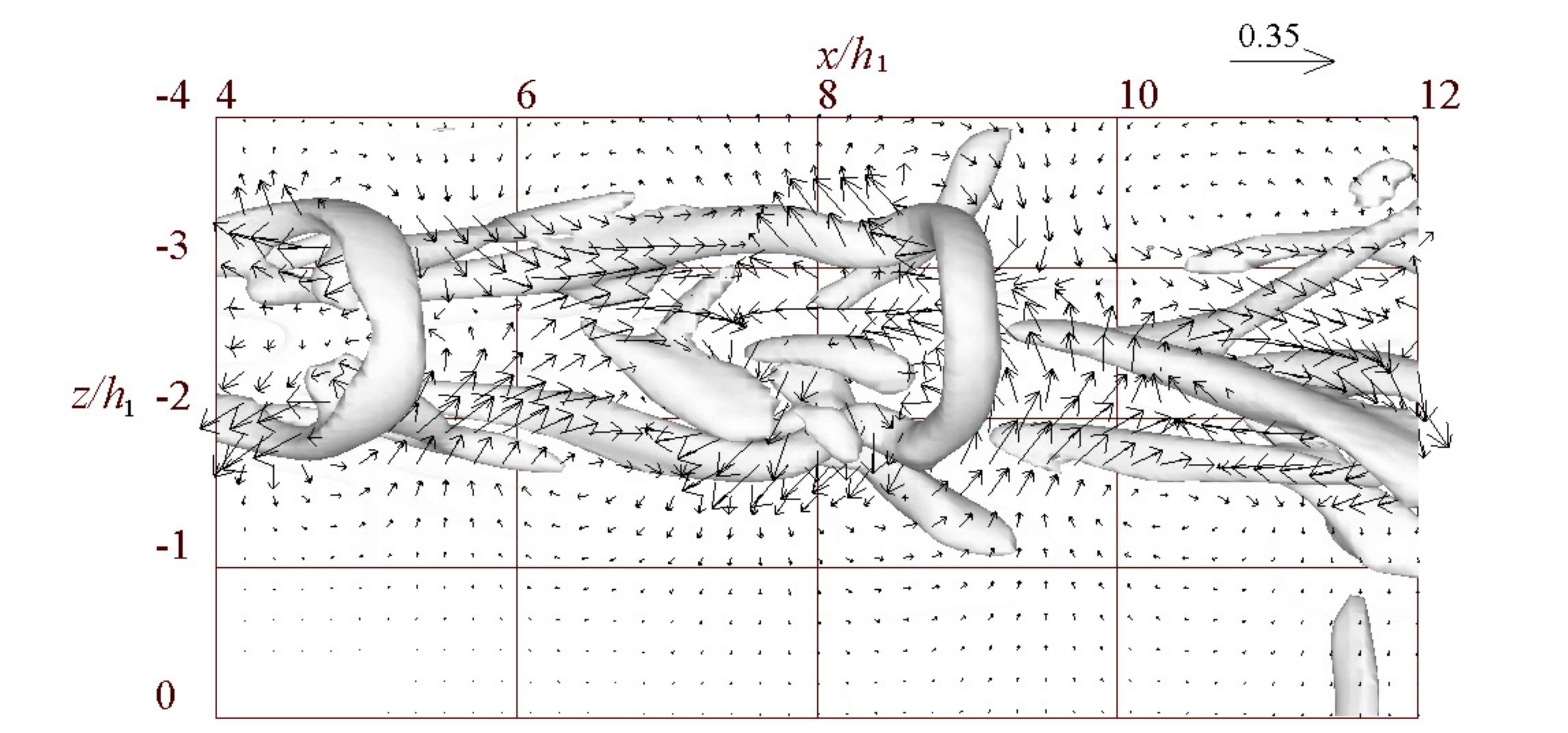} \\
(b) $h_2/h_1=7/8$, $L_z=5h_1$
\end{minipage}
\caption{
Isosurface of curvature of equipressure surface and velocity fluctuation vectors in $x$-$z$ plane at $y/h_1=0.95$: 
Isosurface value is $-7/h_1$.}
\label{curv_vec_xz}
\end{figure}

To clarify the difference between the secondary vortices generated 
between both legs under each condition, 
Fig. \ref{curv_vec_xz} shows the isosurface of curvature of the equipressure surface 
and velocity fluctuation vectors in the $x$-$z$ plane. 
Here, the results at $y/h_1=0.95$ for the cases of $h_2/h_1=1$, $L_z=4h_1$, 
and $h_2/h_1=7/8$, $L_z=5h_1$ 
in which a large difference appears in the secondary vortex are shown. 
For $h_2/h_1=1$ and $L_z= 4h_1$, 
as the hairpin vortices become asymmetric due to the interference between the hairpin vortices, 
the flow due to the Q2 ejection occurs toward the spanwise center side around $x/h_1=4-5$. 
At $x/h_1=5-7$, the asymmetry of the hairpin vortex structure increases 
due to the interference between the hairpin vortices, 
so the velocity fluctuation generated by the Q4 sweep on the spanwise center side increases. 
The collision of these flows generated by the upstream and downstream hairpin vortices 
causes the shear layer to become asymmetric to the center plane of the hill, 
and the secondary vortex seen at $x/h_1=6-8.5$ extends toward the spanwise center side. 
It can be seen that interference occurs between this secondary vortex and the leg 
on the spanwise center side 
and that the fluctuation of the flow to the upstream side is strong around $x/h_1=7$.

On the other hand, for $h_2/h_1=7/8$ and $L_z=5h_1$, 
the interference between the hairpin vortices is weak, 
and the flow asymmetry becomes smaller than that for $h_2/h_1=1$ and $L_z=4h_1$. 
A smaller secondary vortex than that for $h_2/h_1=1$ and $L_z=4h_1$ is generated around $x/h_1=7$. 
Around $x/h_1=11$ downstream, 
two secondary vortices spread above the legs toward the spanwise direction. 
The behavior of this secondary vortex is similar to that of a secondary vortex 
generated around a single hairpin vortex \citep{Yanaoka_et_al_2007b}.

\subsection{Hairpin vortex and heat transport}

\begin{figure}[!t]
\centering
\begin{minipage}{0.49\linewidth}
\centering
\includegraphics[trim=0.1mm 0.5mm 10mm 1.0mm, clip, width=75mm]{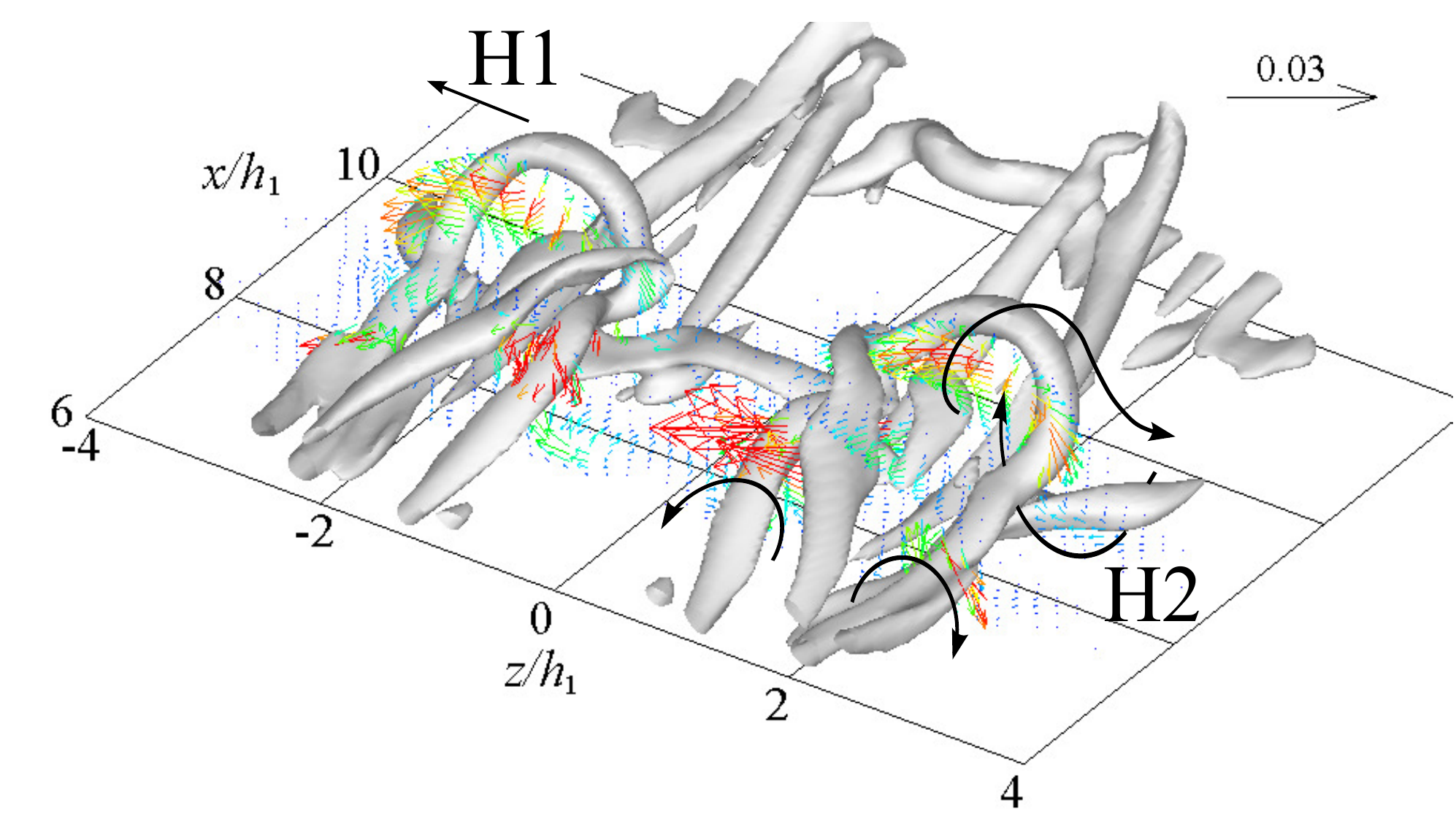} \\
(a)  Perspective view
\end{minipage}
\begin{minipage}{0.49\linewidth}
\centering
\includegraphics[trim=0.1mm 9.5mm 0mm 2mm, clip, width=80mm]{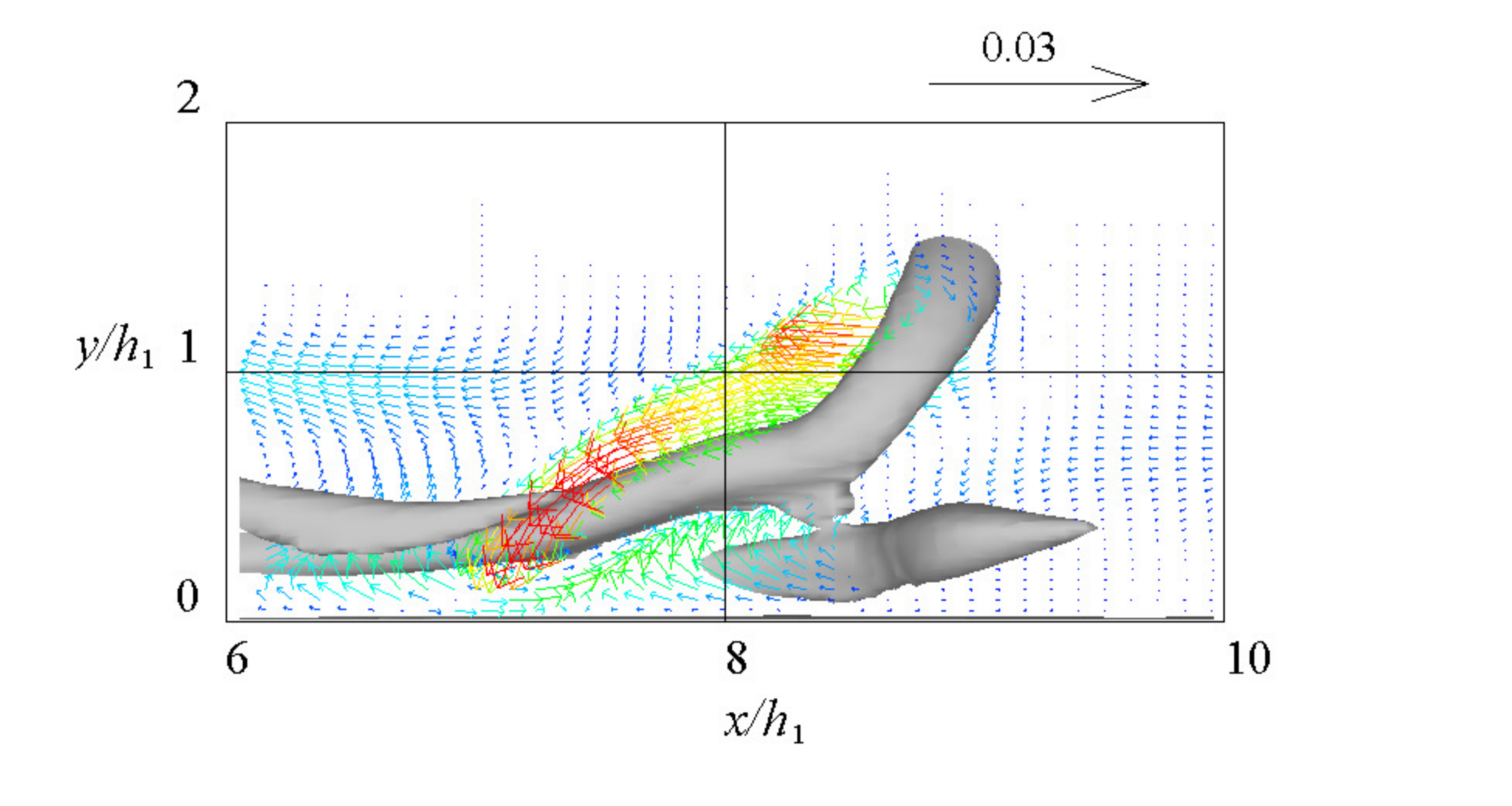} \\
(b) Side view
\end{minipage}
\caption{Isosurface of curvature of equipressure surface 
and turbulent heat flux vectors for $h_2/h_1=1$ and $L_z=4h_1$: 
Isosurface value is $-7/h_1$.}
\label{uvwt}
\end{figure}

To clarify heat transport around the hairpin vortex, 
Fig. \ref{uvwt} shows instantaneous vortex structures and turbulent heat flux vector $u_i'\theta'$ 
for $h_2/h_1=1$ and $L_z=4h_1$. 
Figure \ref{uvwt}(a) shows the heat flux around the head and leg of the hairpin vortex, 
and the arrows in this figure represent the heat flow around the hairpin vortex. 
Figure \ref{uvwt}(b) shows the cross-section at $z/h_1=2.5$, 
where the heat flux outside the hairpin vortex and around the horn-shaped secondary vortex can be observed. 
From Fig. \ref{uvwt}, we can see the magnitude and direction of heat transport 
due to the flow around the hairpin vortex. 
The tendency of heat transport was the same in all conditions, 
so the other results were omitted.

An upstream strong heat flux (H1) due to Q2 ejection around the hairpin vortex 
and heat flux (H2) to the mainstream side due to the legs and horn-shaped secondary vortices are generated, 
and active transport of hot fluid arises. 
Around the legs of the hairpin vortex, 
the high-temperature fluid transported by the ascending flow 
due to the rotation of the legs flows toward the wall surface outside of the legs. 
Around the legs on the spanwise center side in each hairpin vortex, 
because the turbulence increase due to interference between hairpin vortices 
and the flow toward the mainstream side due to secondary vortices occur, 
high-temperature fluid is transported more actively.

\begin{figure}[!t]
\centering
\begin{minipage}{0.48\linewidth}
\centering
\includegraphics[trim=0mm 10mm 0mm 0mm, clip, width=75mm]{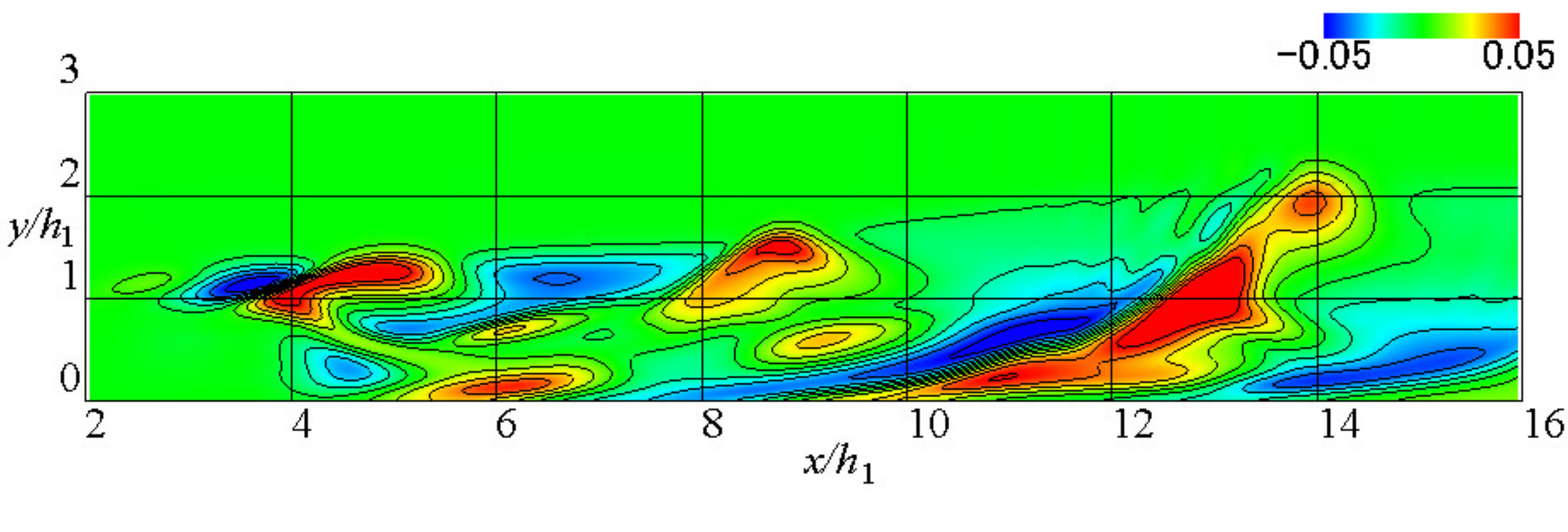} \\
(a) $h_2/h_1=1$, $L_z=4h_1$ ($z/h_1=-2.0$)
\end{minipage}
\begin{minipage}{0.48\linewidth}
\centering
\includegraphics[trim=0mm 10mm 0mm 0mm, clip, width=75mm]{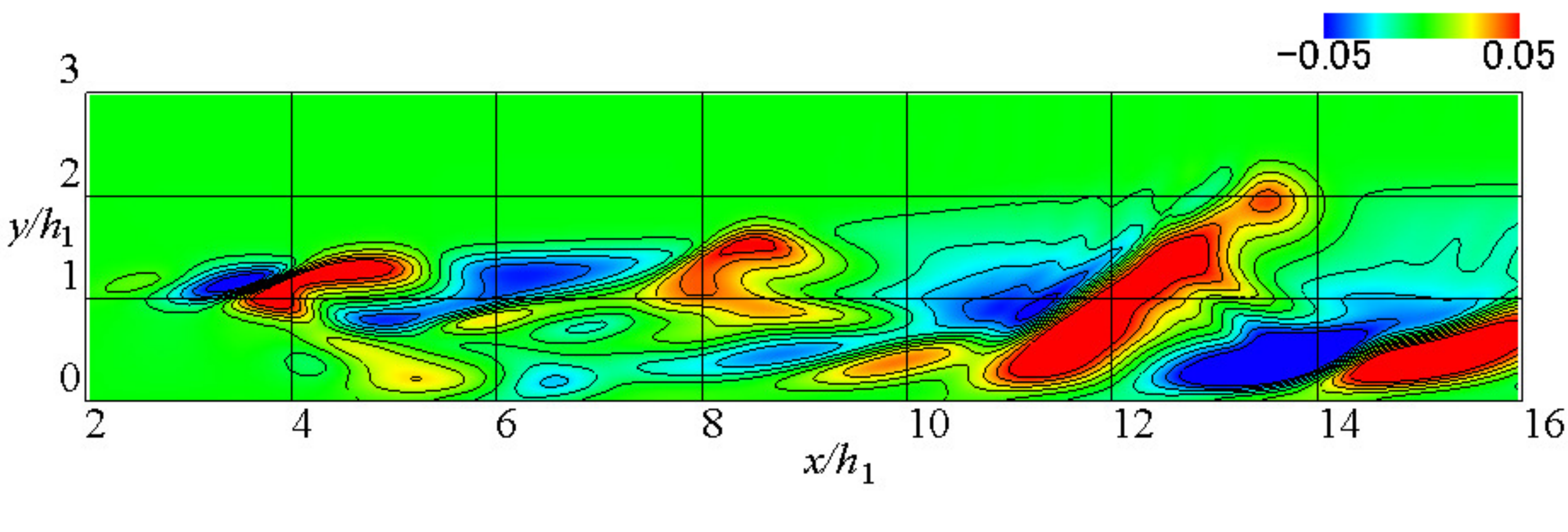} \\
(b) $h_2/h_1=1$, $L_z=5h_1$ ($z/h_1=-2.5$)
\end{minipage}
\vspace*{0.25\baselineskip}
\begin{minipage}{0.48\linewidth}
\centering
\includegraphics[trim=0mm 10mm 0mm 0mm, clip, width=75mm]{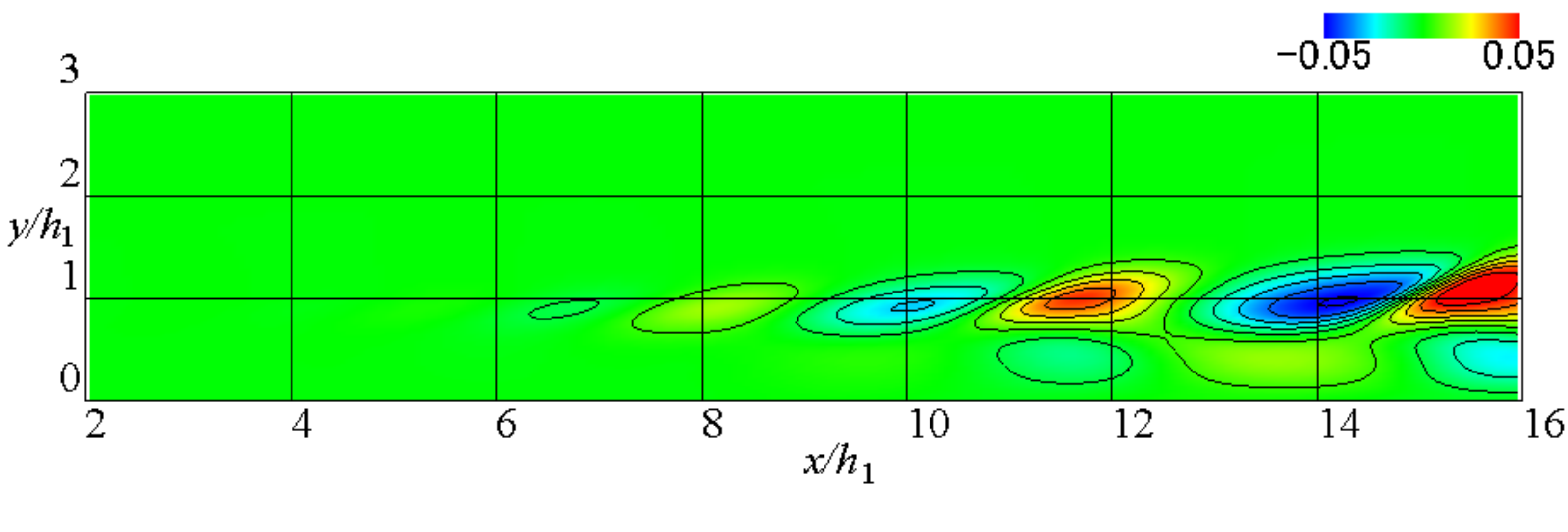} \\
(c) $h_2/h_1=7/8$, $L_z=4h_1$ ($z/h_1=1.4$)
\end{minipage}
\begin{minipage}{0.48\linewidth}
\centering
\includegraphics[trim=0mm 10mm 0mm 0mm, clip, width=75mm]{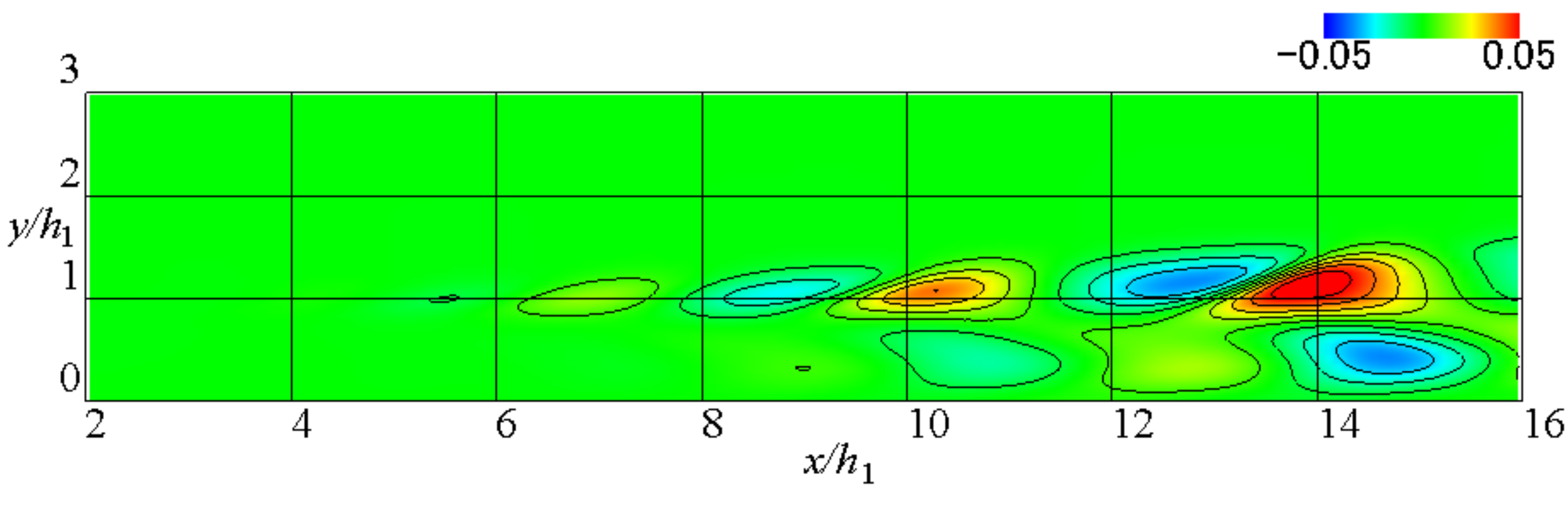} \\
(d) $h_2/h_1=7/8$, $L_z=5h_1$ ($z/h_1=2.2$)
\end{minipage}
\caption{
Contours of temperature fluctuation in $x$-$y$ plane: 
Contour interval is 0.01 from $-0.05$ to 0.05.}
\label{t_side}
\end{figure}

Figure \ref{t_side} shows the distribution of temperature fluctuations 
in the $x$-$y$ plane at the same $z/h_1$ as Fig. \ref{curv_vec}. 
Upstream of the head, 
positive fluctuations occur due to the transport of hot fluid by the Q2 ejection, 
and downstream of the head, 
negative fluctuations occur due to the inflow of cold fluid by the Q4 sweep. 
For $h_2/h_1=1$ and $L_z=4h_1$, compared to the result of $L_z=5h_1$, 
more intensive positive fluctuations are observed in a wide range of $x/h_1=5-7$, 
and $9-13$ near the wall surface, and heat transport becomes active. 
This is because the legs of the hairpin vortex remove the hot fluid 
near the wall surface upward. 
In the case of a single hairpin vortex \citep{Yanaoka_et_al_2007b}, 
no such positive fluctuation was observed over a wide area near the wall surface.

At $h_2/h_1=7/8$, as the rotation around the head is weaker than at $h_2/h_1=1$, 
the region of temperature fluctuation caused by the head becomes smaller. 
For $L_z=5h_1$, the ascending flow by both legs is weak, and for $L_z=4h_1$, 
the rotation of one leg of the hairpin vortex is weak, 
so no strong positive fluctuation occurs near the wall surface.

\begin{figure}[!t]
\centering
\begin{minipage}{0.48\linewidth}
\centering
\includegraphics[trim=0mm 0mm 0mm 30mm, clip, width=65mm]{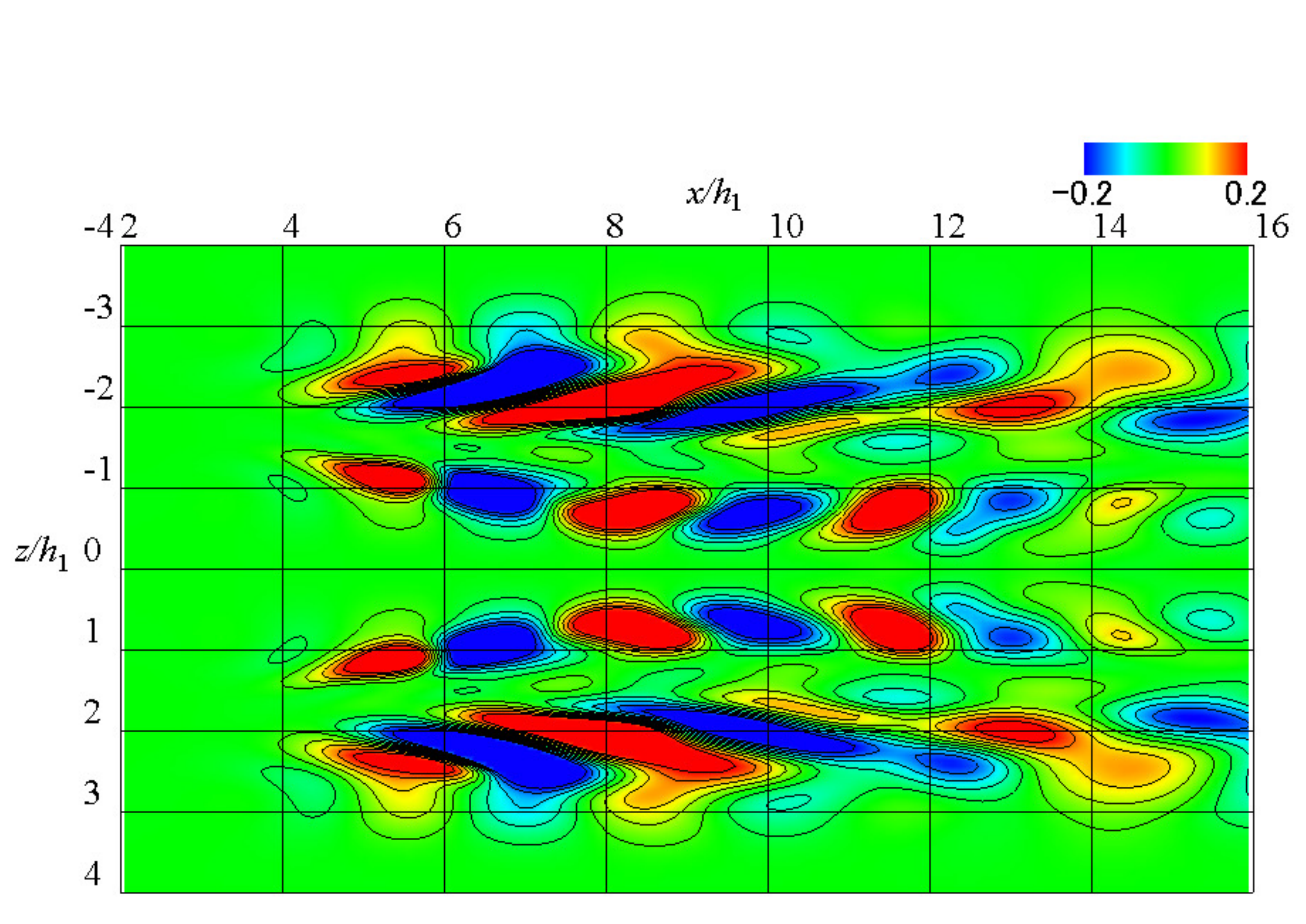} \\
(a) $h_2/h_1=1$, $L_z=4h_1$
\end{minipage}
\begin{minipage}{0.48\linewidth}
\centering
\includegraphics[trim=0mm 0mm 0mm 30mm, clip, width=65mm]{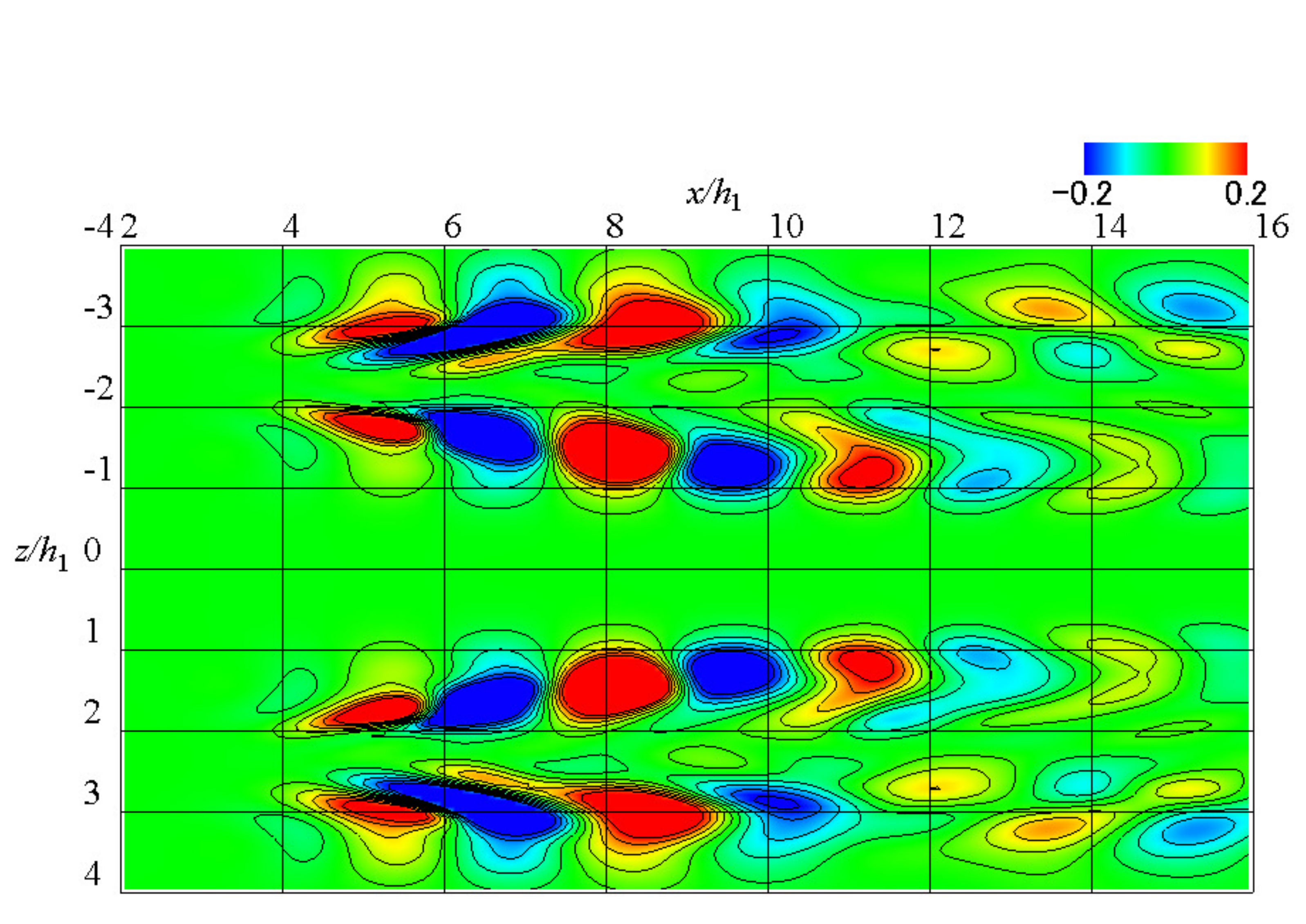} \\
(b) $h_2/h_1=1$, $L_z=5h_1$
\end{minipage}
\vspace*{0.5\baselineskip}
\begin{minipage}{0.48\linewidth}
\centering
\includegraphics[trim=0mm 0mm 0mm 30mm, clip, width=65mm]{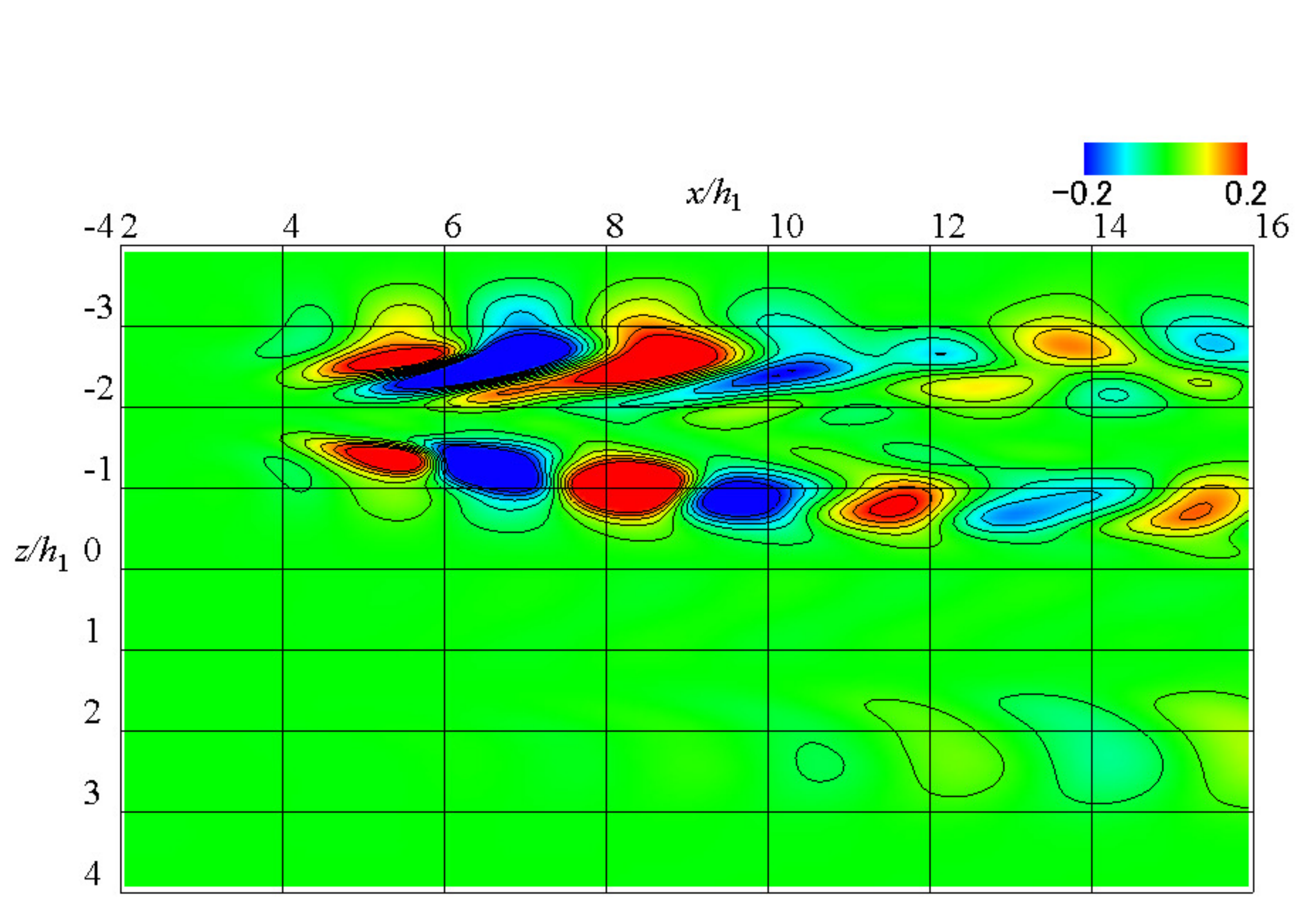} \\
(c) $h_2/h_1=7/8$, $L_z=4h_1$
\end{minipage}
\begin{minipage}{0.48\linewidth}
\centering
\includegraphics[trim=0mm 0mm 0mm 30mm, clip, width=65mm]{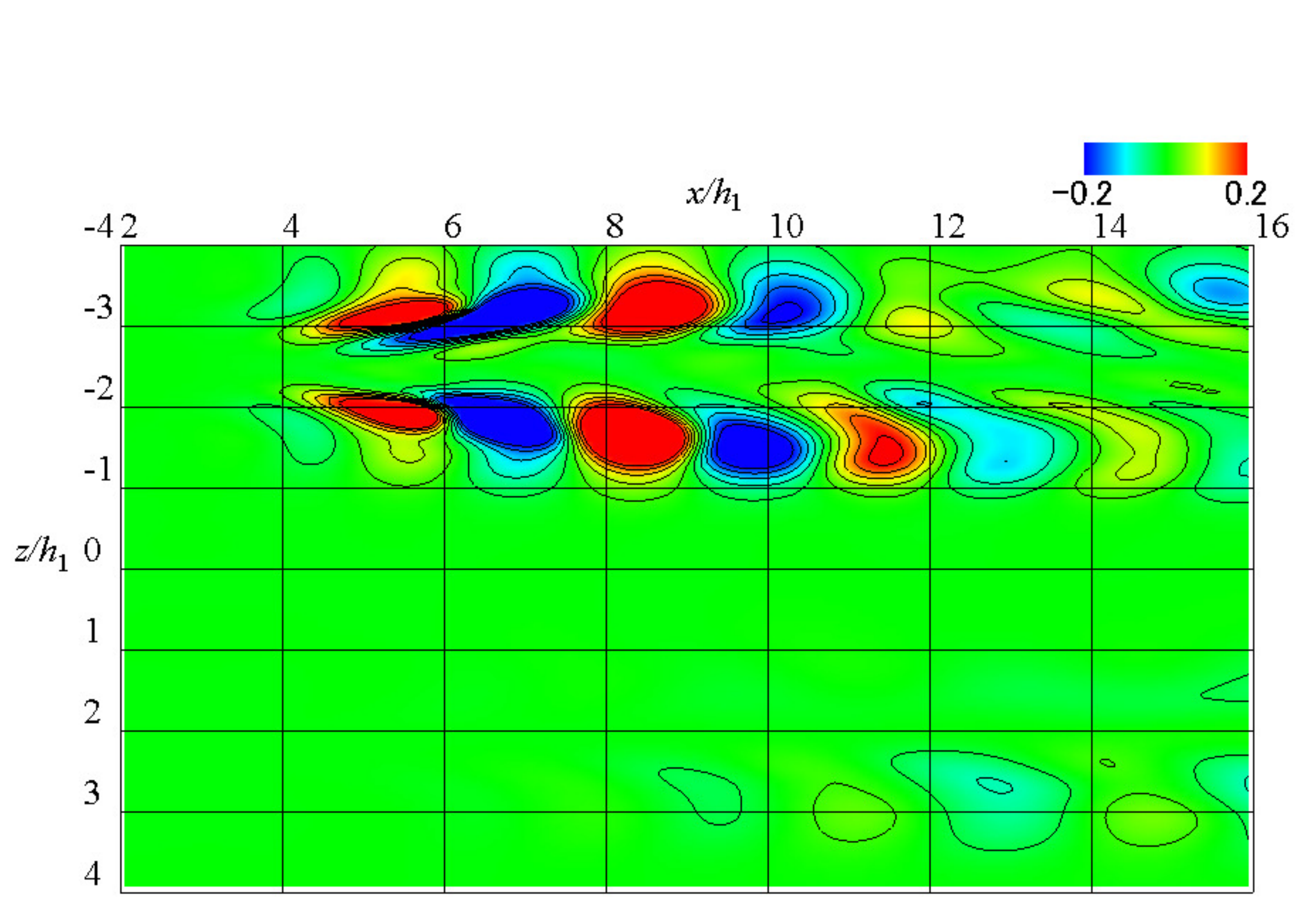} \\
(d) $h_2/h_1=7/8$, $L_z=5h_1$
\end{minipage}
\caption{Contours of Nusselt number fluctuation in $x$-$z$ plane: 
Contour interval is 0.04 from $-0.2$ to 0.2.}
\label{nu_top}
\end{figure}

Next, Fig. \ref{nu_top} shows the fluctuation distribution of the Nusselt number in the $x$-$z$ plane. 
In Fig. \ref{curv_vec}, 
the legs of the hairpin vortices existing at $x/h_1=6-8$ and $10-14$ form the ascending flow. 
Furthermore, as shown in Fig. \ref{uvwt}, 
the legs and horn-shaped secondary vortices generate heat flux 
toward the mainstream side and transport the high-temperature fluid. 
Such a flow enhances heat transfer. 
Therefore, for all conditions, 
positive fluctuations occur around $x/h_1=8$ and 13 behind the hill at $z/h_1<0$. 
Negative fluctuations arise around $x/h_1=7$, 10, and 13. 
This is because, as seen in Fig. \ref{uvwt}, 
the high-temperature fluid transported by the legs flows toward the wall surface 
by the downward flow generated outside the legs. 
As a result, the heat transfer coefficient decreases. 
In the case of a single hairpin vortex \citep{Yanaoka_et_al_2009}, 
the fluctuation distribution was symmetric to the central cross-section of the hill, 
but it is asymmetric in this study.

In the case of the same $L_z$, for $h_2/h_1=1$, 
as stronger interference between hairpin vortices occurs compared to $h_2/h_1=7/8$, 
high fluctuations appear on the spanwise center side. 
In each $h_2/h_1$, when the hill spacing is narrowed, 
the interference between hairpin vortices becomes more intensive, 
so high fluctuations occur in the center of the spanwise direction, even downstream. 
In addition, the asymmetry of the hairpin vortex increases 
due to the intensive interference between hairpin vortices, 
and most of one leg at $z/h_1<-2$ in the hairpin vortex approach the vicinity of the wall surface. 
As a result, the influence of the legs is widespread, 
and the positive fluctuation region around $z/h_1=-2$ near $x/h_1=8$ and 13 increases. 
It will be described later that the turbulence near the wall surface increases 
due to the closeness of the legs.

For $h_2/h_1=7/8$, a weak fluctuation occurs behind the hill at $z/h_1 \ge 0$ 
due to a small-scale hairpin vortex. 
Around the small hairpin vortex, 
the ascending flow by the head and legs is weaker than that of the large-scale hairpin vortex. 
Therefore, heat transport decays and intensive fluctuations 
in heat transfer coefficient do not occur.

\subsection{Mean properties}

\begin{figure}[!t]
\begin{minipage}{0.48\linewidth}
\centering
\includegraphics[trim=0mm 2.1mm 0mm 4.2mm, clip, width=88mm]{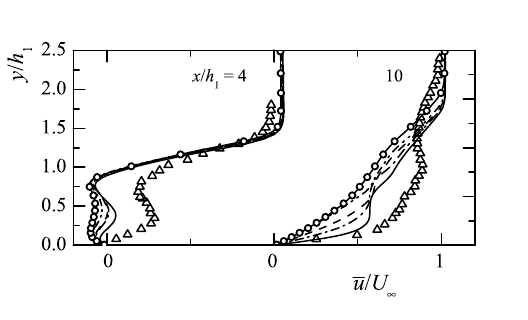}
\caption{Time-averaged streamwise velocity distributions: 
\mbox{---,} $h_2/h_1=1$, $L_z=4h_1$;  
\mbox{- - -,} $h_2/h_1=1$, $L_z=5h_1$; 
\mbox{-- $\cdot$ --,} $h_2/h_1=7/8$, $L_z=4h_1$; 
\mbox{-- $\cdot$ $\cdot$ --,} $h_2/h_1=7/8$, $L_z=5h_1$; 
--$\circ$--, Yanaoka et al. ($Re=500$);
$\bigtriangleup$, Acarlar--Smith ($Re=750$).}
\label{um}
\end{minipage}
\hspace{4.3mm}
\begin{minipage}{0.49\linewidth}
\vspace*{-1.1\baselineskip}
\begin{minipage}{0.49\linewidth}
\centering
\includegraphics[trim=0mm 2.1mm 0mm 4.2mm, clip, width=38mm]{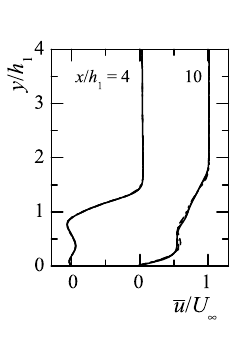} \\
(a) $h_2/h_1=1$, $L_z=4h_1$
\end{minipage}
\vspace*{0.5\baselineskip}
\begin{minipage}{0.49\linewidth}
\centering
\includegraphics[trim=0mm 2.1mm 0mm 4.1mm, clip, width=38mm]{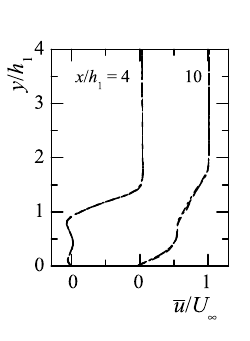} \\
(b) $h_2/h_1=7/8$, $L_z=4h_1$
\end{minipage}
\caption{Time-averaged streamwise velocity distributions: 
-- $\cdot$ $\cdot$ --, grid1; -- $\cdot$ --, grid2; - - -, grid3; ---, grid4.}
\label{um_hikaku}
\end{minipage}
\end{figure}

Figure \ref{um} shows the time-averaged streamwise velocity distribution $\overline{u}$ 
for each condition. 
These distributions are the values in the central cross-section of the hill at $z/h_1<0$. 
All distributions are compared with the experimental values at $Re=750$ by \citet{Acarlar&Smith_1987a} 
and the calculations by the authors \citep{Yanaoka_et_al_2007b} for a single hairpin vortex. 
In the distribution of $\overline{u}$ at $x/h_1=4$, 
the result of \citet{Acarlar&Smith_1987a} shows no reverse flow. 
Conversely, the result of the authors \citep{Yanaoka_et_al_2007b} 
and this calculation show reverse flow in all distributions. 
At $x/h_1=10$, $\overline{u}$ near the wall surface for $L_z=4h_1$ 
is faster than the result for a single hairpin vortex \citep{Yanaoka_et_al_2007b} 
regardless of $h_2/h_1$. 
In this study, the hairpin vortices become asymmetric 
due to the interference between the hairpin vortices. 
As a result, the leg of the hairpin vortex approaches 
the central cross-section of the hill, 
increasing $\overline{u}$ near the wall surface. 
On the other hand, the result for $h_2/h_1=7/8$ and $L_z=5h_1$ show 
almost the same velocity as that for a single hairpin vortex, 
indicating effects of interference between hairpin vortices are weak.

Figure \ref{um_hikaku} shows the distribution of $\overline{u}$ 
obtained by the four grids for $h_2/h_1=1$ and 7/8 at $L_z=4h_1$. 
These distributions are the values in the central cross-section of the hill at $z/h_1<0$. 
We can confirm there is no intensive difference in the results 
depending on the number of grid points.

\begin{figure}[!t]
\centering
\includegraphics[trim=4.5mm 0.9mm 0mm 2mm, clip, width=75mm]{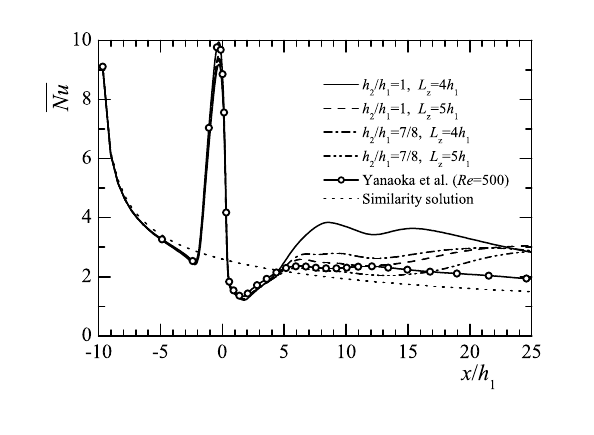}
\vspace*{-1.5\baselineskip}
\caption{Time-averaged Nusselt number distributions.}
\label{nu}
\end{figure}

To investigate the effect of hairpin vortices on heat transfer, 
Fig. \ref{nu} shows the time-averaged Nusselt number $\overline{Nu}$ 
in the central cross-section of the hill at $z/h_1<0$. 
All distributions are compared with the calculation result 
for a single hairpin vortex \citep{Yanaoka_et_al_2007b} 
and the value obtained from the similarity solution of the laminar boundary layer 
without flow separation \citep{Lighthill_1950}. 
The similarity solution is given as 
$Nu={Nu_0}/{[1-(s/x)]^{1/3}}$ and $Nu_0=0.464Pr^{1/3}Re^{1/2}_{x}$, 
where $s$ is the distance from the start point 
where the laminar boundary layer develops to the start point of heating, 
and in this study, $s=10h_1$.

In all conditions, $\overline{Nu}$ shows a lower value in the recirculation region 
than the similarity solution. 
Downstream from $x/h_1=5$, $\overline{Nu}$ is higher than the similarity solution 
because the hairpin vortex increases heat transport. 
Downstream from $x/h_1=5$ at $h_2/h_1=1$ and 7/8 for $L_z=4h_1$, 
and downstream from $x/h_1=12$ and 14 at $h_2/h_1=1$ and 7/8 for $L_z=5h_1$, respectively, 
$\overline{Nu}$ is higher than the result for a single hairpin vortex. 
In this study, the hairpin vortex becomes asymmetric 
due to the interference between hairpin vortices. 
As a result, $\overline{Nu}$ increases 
because the leg and the horn-shaped secondary vortex pass 
near the central cross-section of the hill 
and remove the high-temperature fluid near the wall surface. 
This tendency is most prominent for $h_2/h_1=1$ and $L_z=4h_1$, 
so $\overline{Nu}$ at this condition is the highest among all results.

\subsection{Turbulence characteristics}

\begin{figure}[!t]
\centering
\begin{minipage}{0.48\linewidth}
\centering
\includegraphics[trim=0mm 3mm 0mm 0mm, clip, width=65mm]{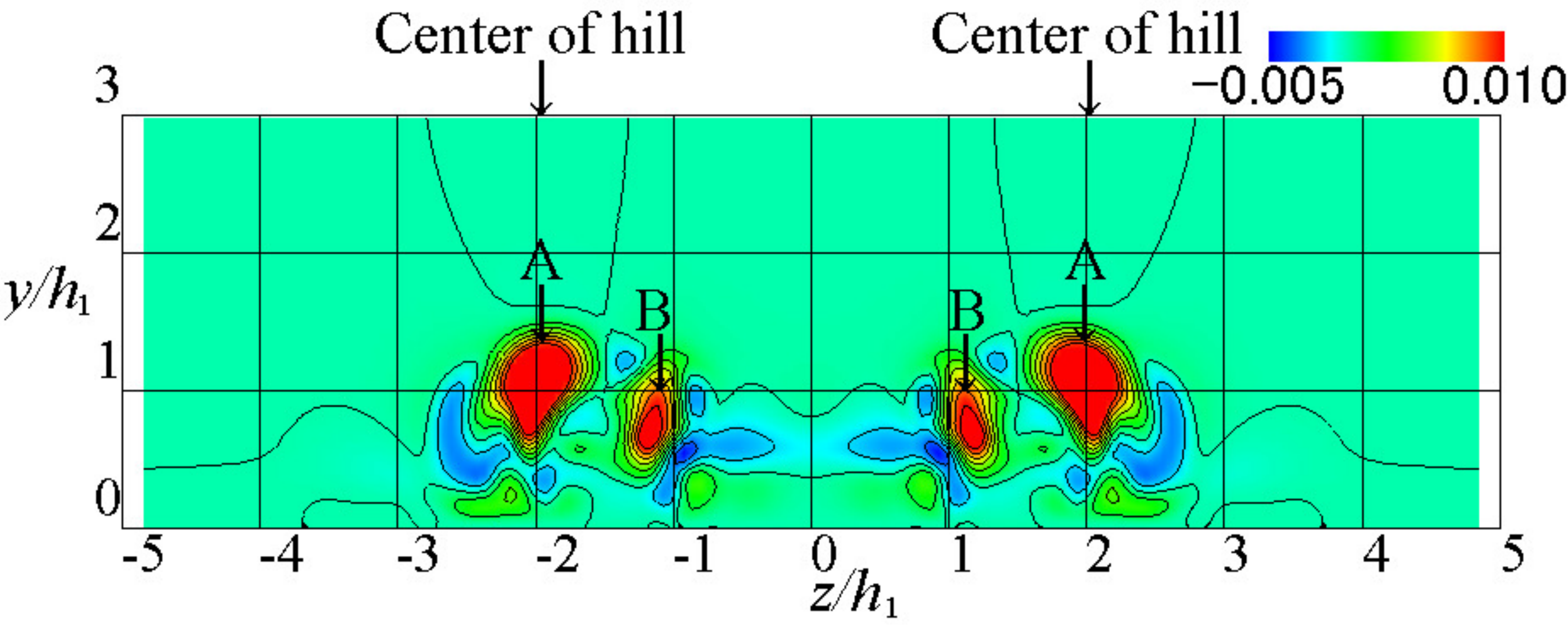} \\
(a) $h_2/h_1=1$, $L_z=4h_1$ \\
\end{minipage}
\begin{minipage}{0.48\linewidth}
\centering
\includegraphics[trim=0mm 3mm 0mm 0mm, clip, width=65mm]{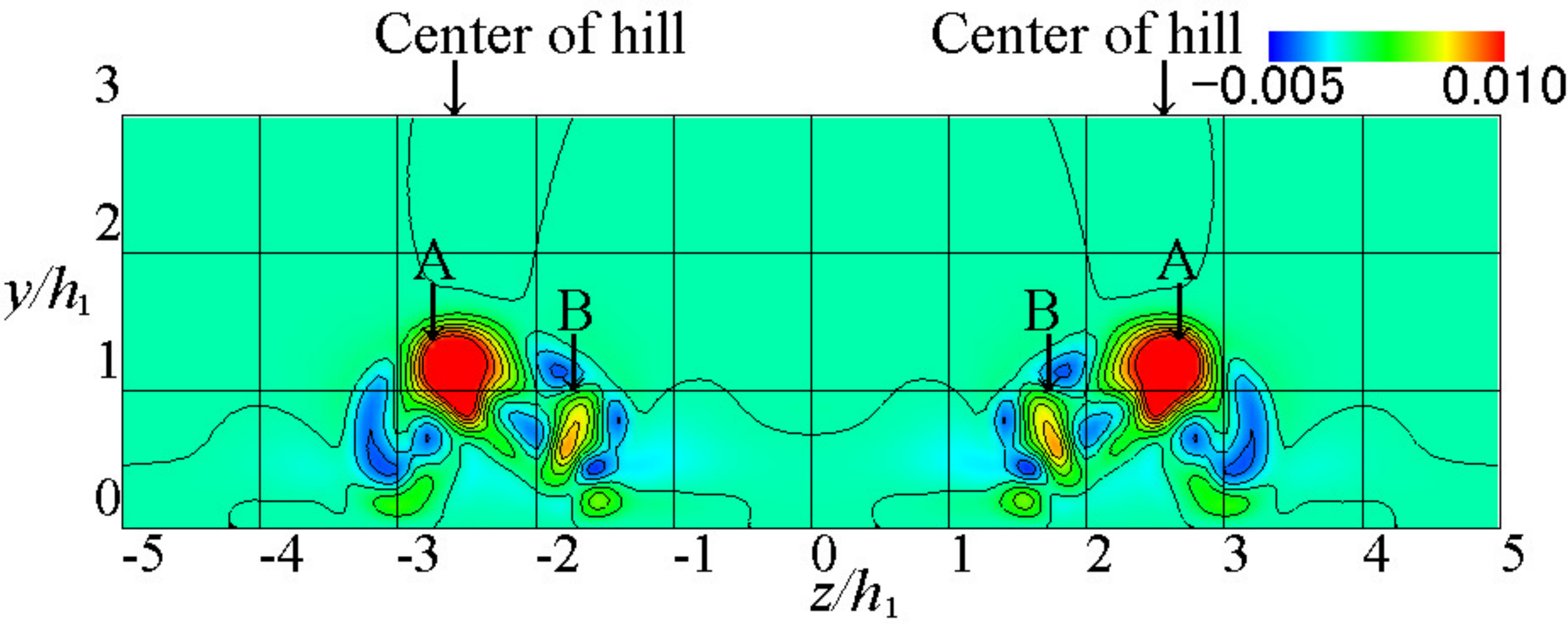} \\
(b) $h_2/h_1=1$, $L_z=5h_1$ \\
\end{minipage}
\vspace*{0.2\baselineskip}
\begin{minipage}{0.48\linewidth}
\centering
\vspace*{0.5\baselineskip}
\includegraphics[trim=0mm 3mm 0mm 0mm, clip, width=65mm]{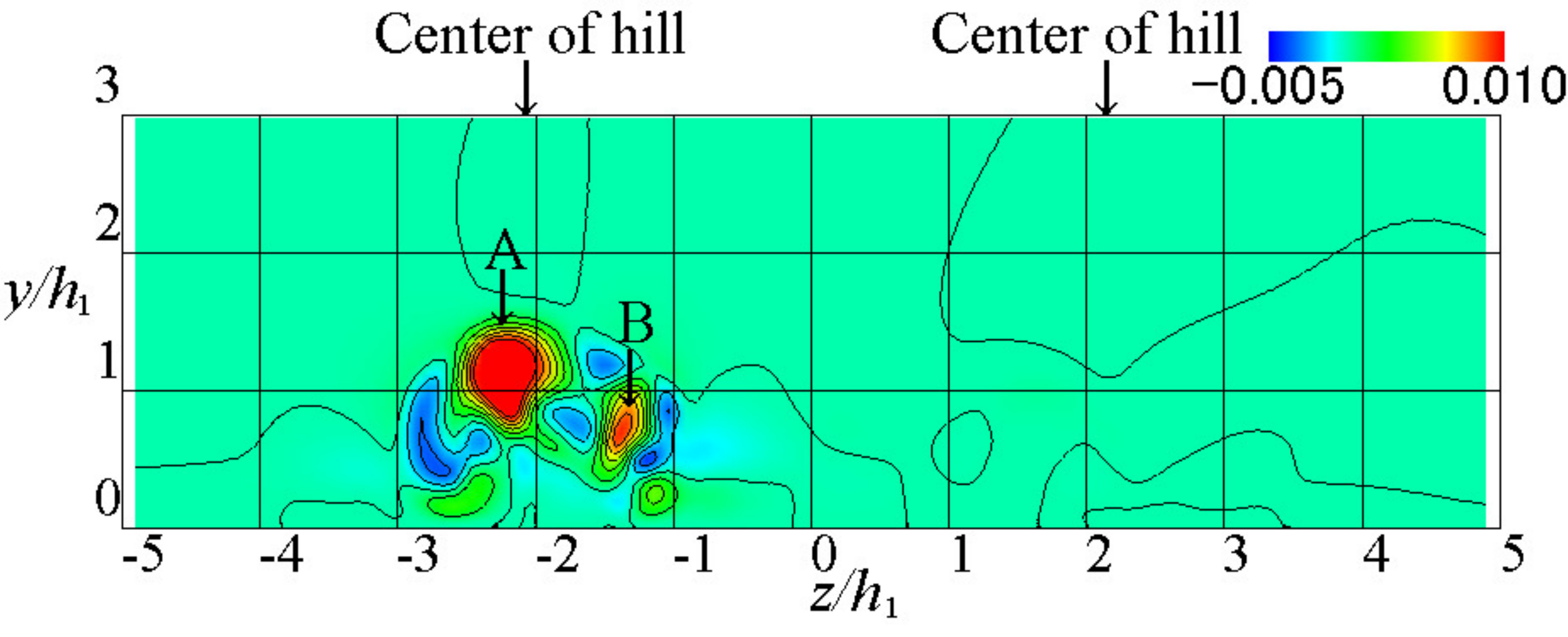} \\
(c) $h_2/h_1=7/8$, $L_z=4h_1$ \\
\end{minipage}
\begin{minipage}{0.48\linewidth}
\centering
\vspace*{0.5\baselineskip}
\includegraphics[trim=0mm 3mm 0mm 0mm, clip, width=65mm]{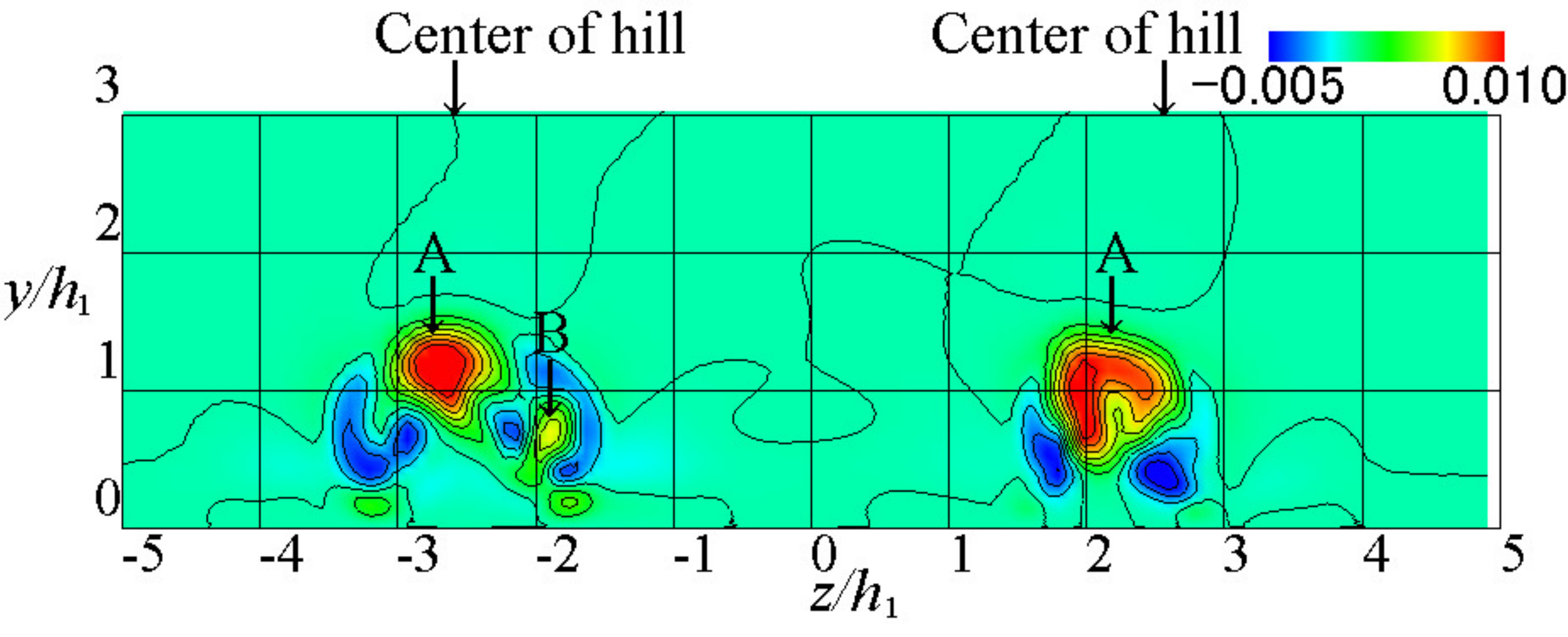} \\
(d) $h_2/h_1=7/8$, $L_z=5h_1$ \\
\end{minipage}
\caption{
Contours of Reynolds shear stress $-\overline{u'v'}$ in $y$-$z$ plane at $x/h_1=8.0$: 
Contour interval is 0.0015 from $-0.005$ to 0.010.}
\label{uv_x8}
\end{figure}

\begin{figure}[!t]
\centering
\begin{minipage}{0.48\linewidth}
\centering
\includegraphics[trim=0mm 3mm 0mm 0mm, clip, width=65mm]{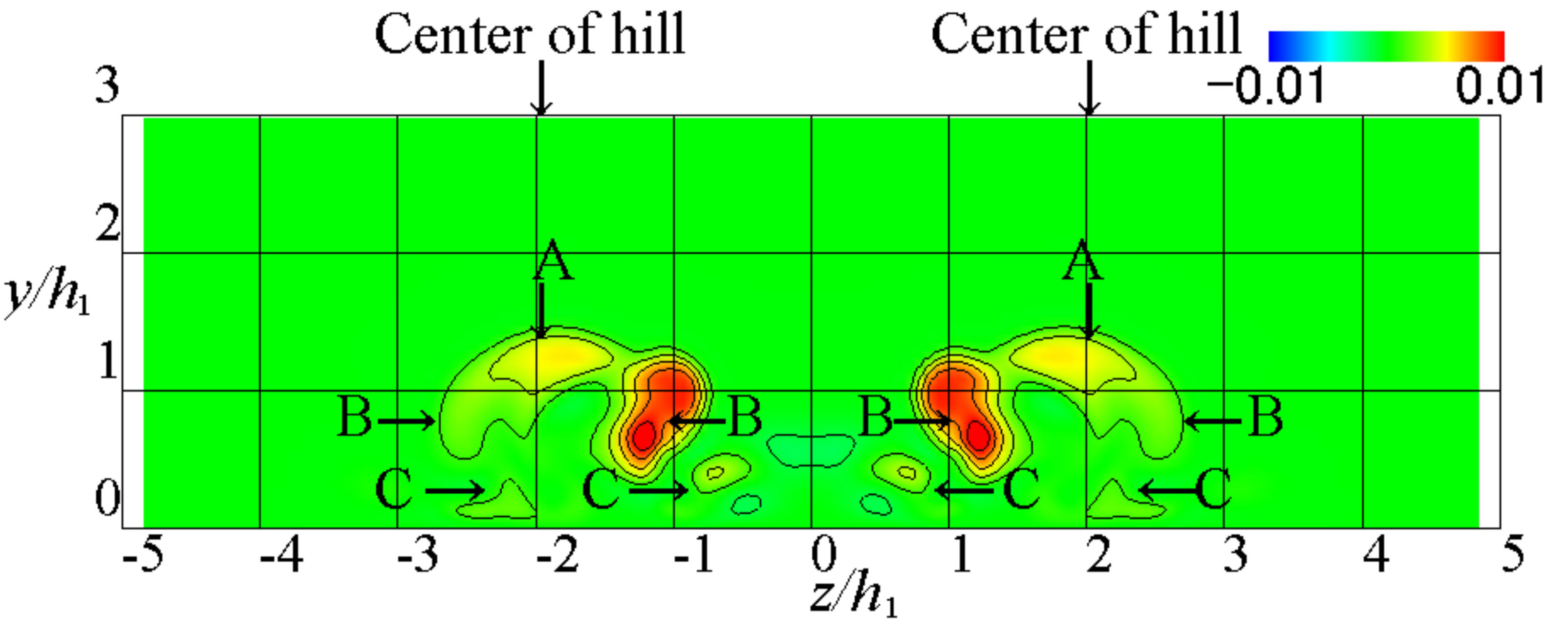} \\
(a) $h_2/h_1=1$, $L_z=4h_1$ \\
\end{minipage}
\begin{minipage}{0.48\linewidth}
\centering
\includegraphics[trim=0mm 3mm 0mm 0mm, clip, width=65mm]{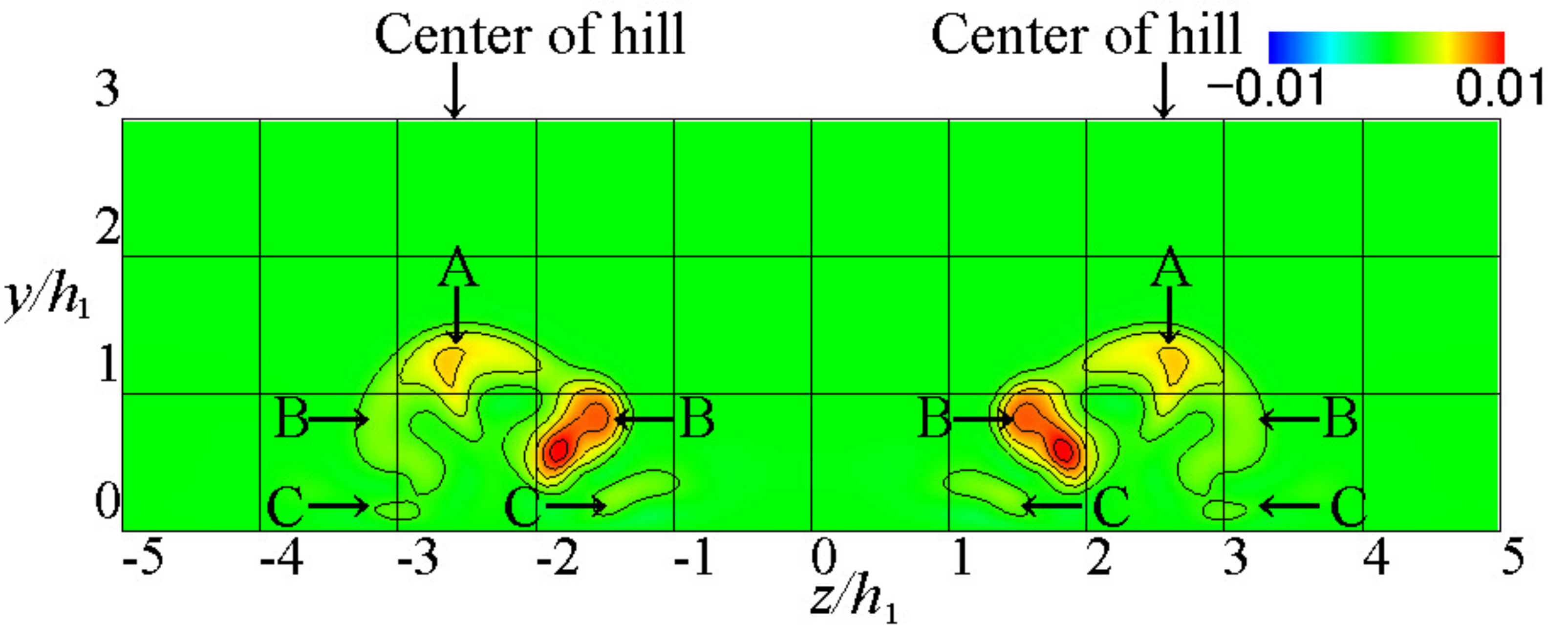} \\
(b) $h_2/h_1=1$, $L_z=5h_1$ \\
\end{minipage}
\begin{minipage}{0.48\linewidth}
\centering
\includegraphics[trim=0mm 3mm 0mm 0mm, clip, width=65mm]{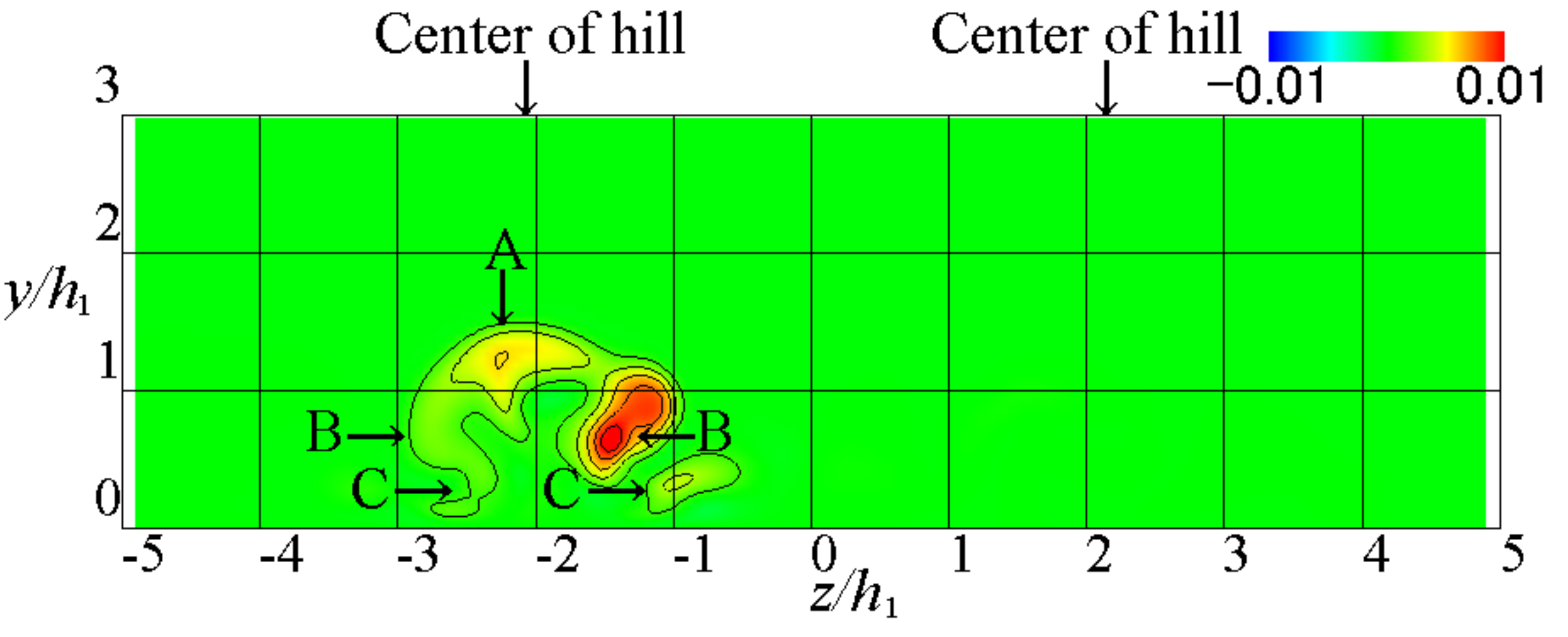} \\
(c) $h_2/h_1=7/8$, $L_z=4h_1$ \\
\end{minipage}
\begin{minipage}{0.48\linewidth}
\centering
\includegraphics[trim=0mm 3mm 0mm 0mm, clip, width=65mm]{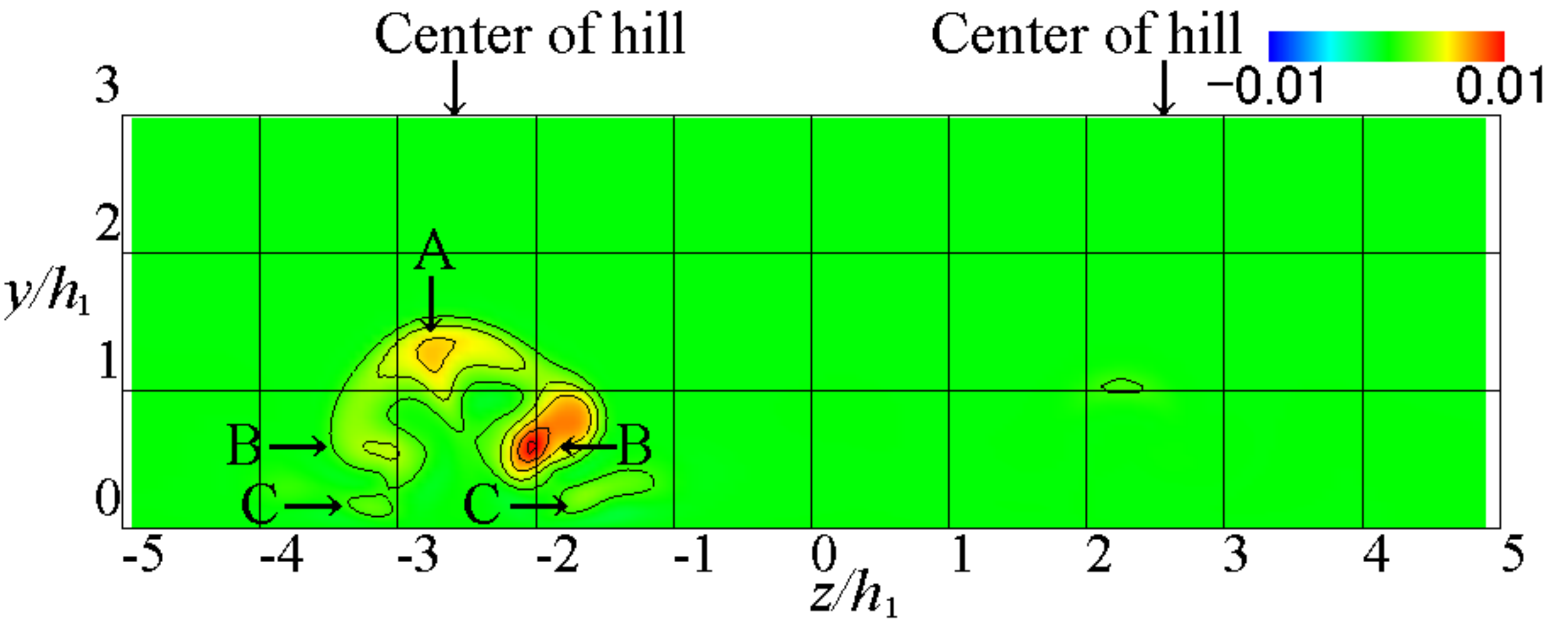} \\
(d) $h_2/h_1=7/8$, $L_z=5h_1$ \\
\end{minipage}
\caption{
Contours of turbulent heat flux $-\overline{u'\theta'}$ in $y$-$z$ plane at $x/h_1=8.0$: 
Contour interval is 0.002 from $-0.01$ to 0.01.}
\label{ut_x8}
\end{figure}

\begin{figure}[!t]
\centering
\begin{minipage}{0.48\linewidth}
\centering
\includegraphics[trim=0mm 3mm 0mm 0mm, clip, width=65mm]{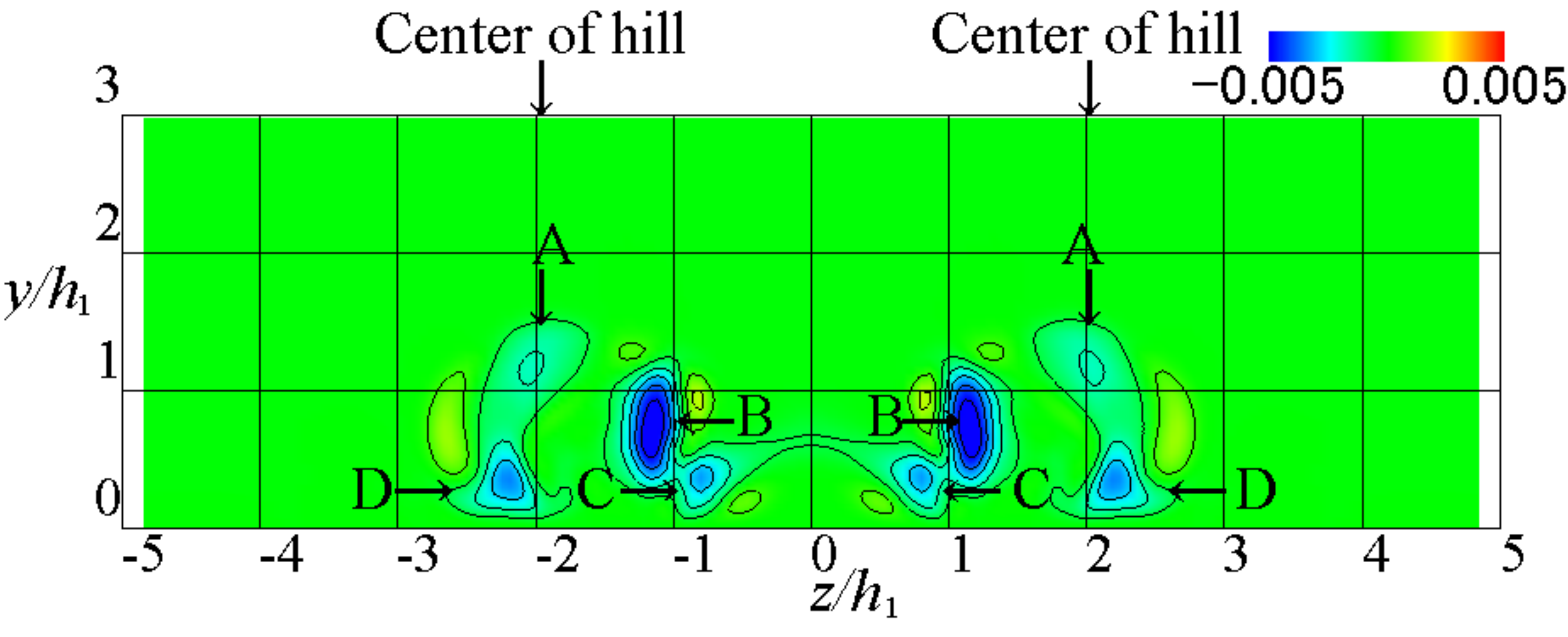} \\
(a) $h_2/h_1=1$, $L_z=4h_1$ \\
\end{minipage}
\begin{minipage}{0.48\linewidth}
\centering
\includegraphics[trim=0mm 3mm 0mm 0mm, clip, width=65mm]{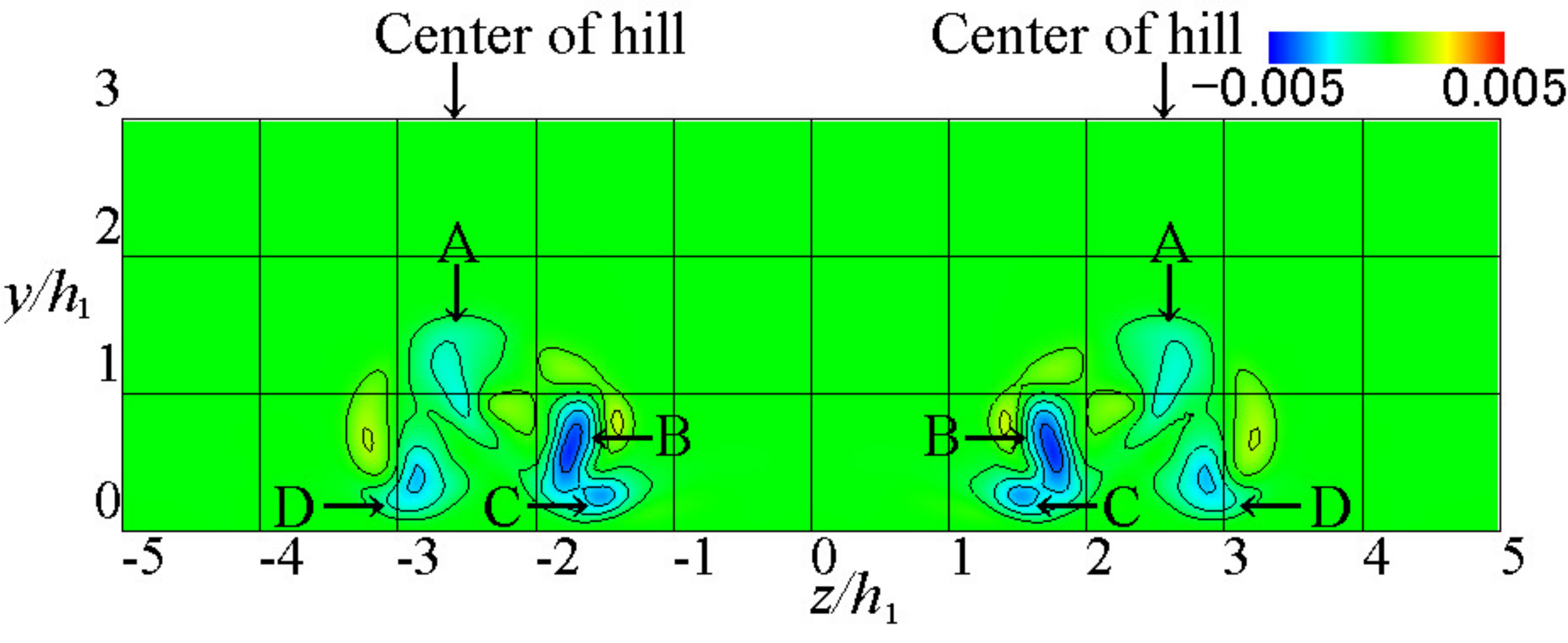} \\
(b) $h_2/h_1=1$, $L_z=5h_1$ \\
\end{minipage}
\vspace*{0.5\baselineskip}
\begin{minipage}{0.48\linewidth}
\centering
\includegraphics[trim=0mm 3mm 0mm 0mm, clip, width=65mm]{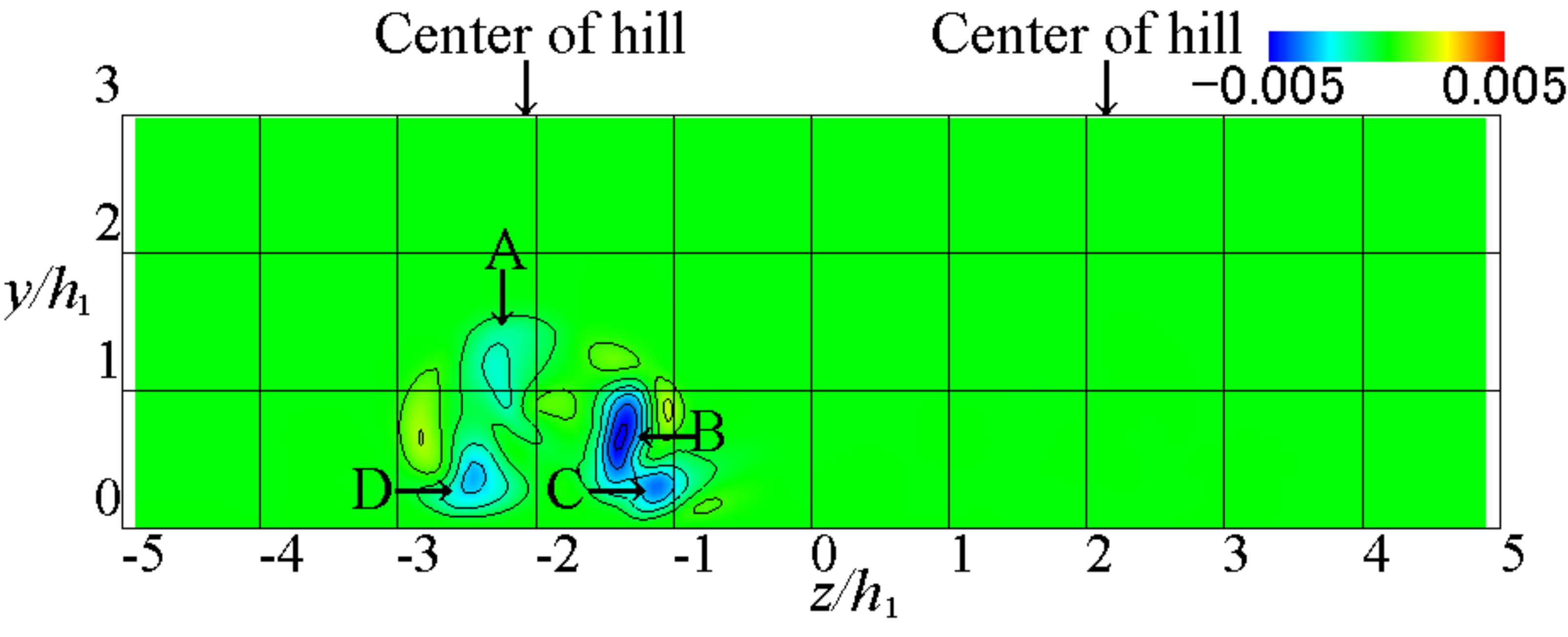} \\
(c) $h_2/h_1=7/8$, $L_z=4h_1$ \\
\end{minipage}
\begin{minipage}{0.48\linewidth}
\centering
\includegraphics[trim=0mm 3mm 0mm 0mm, clip, width=65mm]{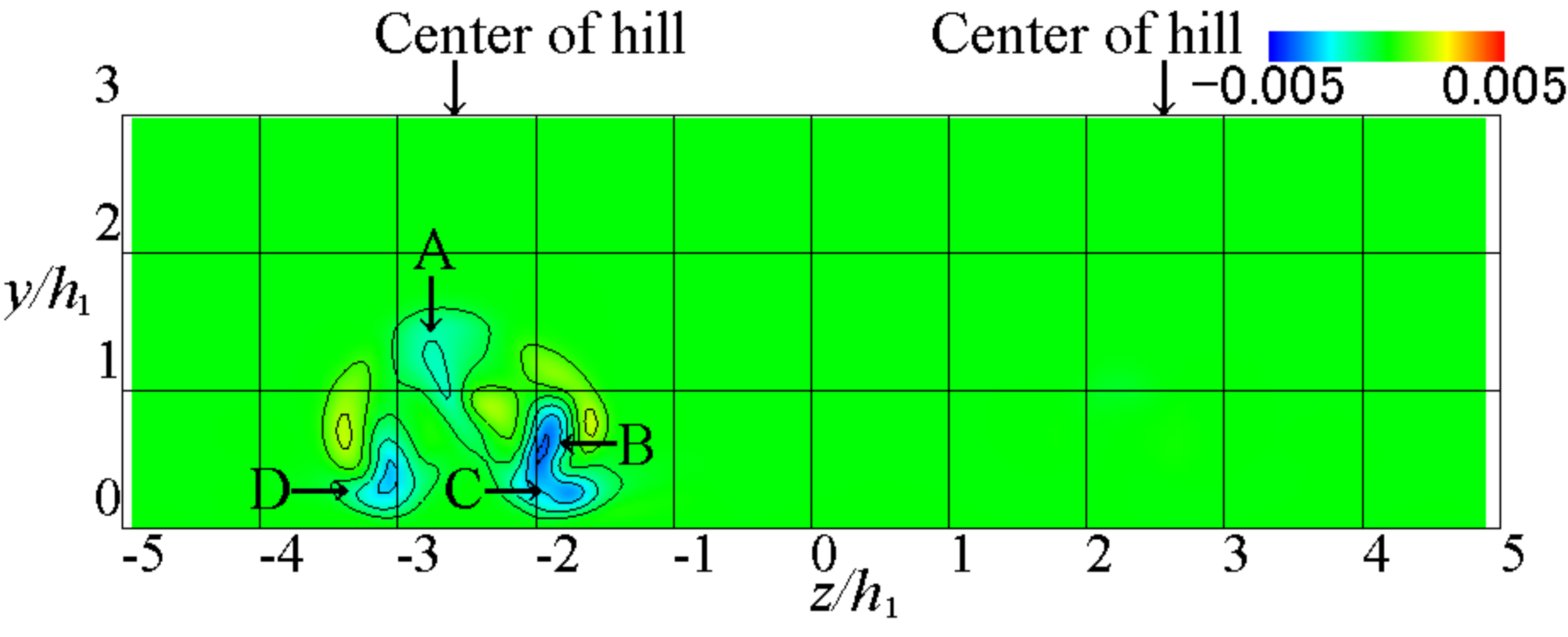} \\
(d) $h_2/h_1=7/8$, $L_z=5h_1$ \\
\end{minipage}
\caption{
Contours of turbulent heat flux $-\overline{v'\theta'}$ in $y$-$z$ plane at $x/h_1=8.0$: 
Contour interval is 0.001 from $-0.005$ to 0.005.}
\label{vt_x8}
\end{figure}

The distributions of Reynolds stress $-\overline{u'v'}$ 
and turbulent heat flux $-\overline{u' \theta'}$ and $-\overline{v'\theta'}$ 
in the $y$-$z$ plane at $x/h_1=8$ are shown in Fig. \ref{uv_x8} to Fig. \ref{vt_x8}, respectively. 
Herein, for each region where a high turbulence region clearly appears, 
the turbulence region caused by the head of the hairpin vortex is labeled A. 
Other turbulence regions are labeled B for the leg and secondary vortex, 
C for the horn-shaped secondary vortex, 
and D for the horn-shaped secondary vortex and leg of the hairpin vortex. 
There are high turbulence regions due to the hairpin vortex behind the hill 
at $z/h_1<0$ for all distributions. 
The turbulence distribution is asymmetric to the center cross-section of the hill, 
unlike the distribution for a single hairpin vortex \citep{Yanaoka_et_al_2007b}. 
This is because vortex structures become asymmetric 
to the center cross-section of the hill due to the interference between hairpin vortices. 
In turbulence region A caused by the head of the hairpin vortex, 
there is no significant increase in turbulence due to changes 
in the scale ratio of the hills or the hill spacing.

In all distributions, as interference between hairpin vortices occurs, 
the turbulence (region B) due to the leg and secondary vortex 
and turbulence (region C) due to the horn-shaped secondary vortex are high, 
and heat transport is active in regions B and C. 
This tendency appears strongly when the hill spacing is narrowed, 
regardless of $h_2/h_1$, and is most remarkable at $h_2/h_1=1$ and $L_z=4h_1$. 
Furthermore, for $h_2/h_1=1$ and $L_z=4h_1$, 
a high turbulence region extends to the spanwise center 
because the horn-shaped secondary vortices are connected. 
For $h_2/h_1=1$ and $L_z=5h_1$, 
the horn-shaped secondary vortices are not coupled at $x/h_1=8$, 
so the high turbulence region does not extend to the spanwise center. 
At $h_2/h_1=7/8$, it can be seen that the turbulence due to the small-scale hairpin vortex is low, 
and the heat transport decays behind the hill at $z/h_1 \ge 0$.

When the hill spacing is narrowed, 
the turbulence region C due to the horn-shaped secondary vortex in $-\overline{u'\theta'}$ 
and the turbulence region D due to the horn-shaped secondary vortex and legs 
in $-\overline{v'\theta}$ approach the central cross-section of the hill 
because the asymmetry of vortex structures increases. 
Therefore, around the central cross-section of the hill, 
heat transport becomes active due to the horn-shaped secondary vortex 
and leg of the hairpin vortex.

\begin{figure}[!t]
\begin{minipage}{0.48\linewidth}
\centering
\includegraphics[trim=2.5mm 6mm 18mm 3mm, clip, width=75mm]{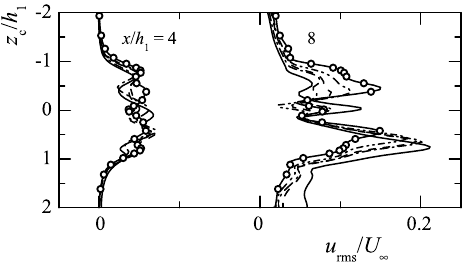} \\
(a) $u_\mathrm{rms}/U_{\infty}$ \\
\end{minipage}
\begin{minipage}{0.48\linewidth}
\centering
\includegraphics[trim=2.5mm 6mm 18mm 3mm, clip, width=75mm]{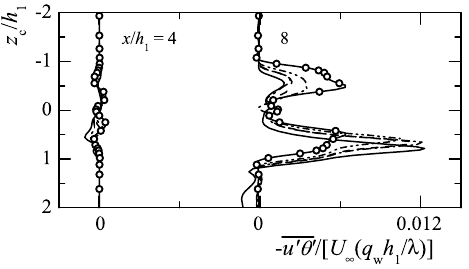} \\
(b) $-\overline{u'\theta'}$/$[U_{\infty}(q_\mathrm{w}h_{1}/\lambda)]$ \\
\end{minipage}
\caption{Turbulence intensity distributions of streamwise velocity fluctuation 
and distribution of turbulent heat flux: 
---, $h_2/h_1=1$, $L_z=4h_1$;  
\mbox{- - -,} $h_2/h_1=1$, $L_z=5h_1$; 
-- $\cdot$ --, $h_2/h_1=7/8$, $L_z=4h_1$; 
-- $\cdot$ $\cdot$ --, $h_2/h_1=7/8$, $L_z=5h_1$; 
--$\circ$--, Yanaoka et al. ($Re=500$).}
\label{rms_xz}
\end{figure}

Figure \ref{rms_xz} shows the distributions in the $z$-direction 
for the rms value $u_\mathrm{rms}$ of the streamwise velocity fluctuation 
and the turbulent heat flux $-\overline{u'\theta'}$. 
The positions in the $y$-direction are $y/h_1=0.65$ and 0.6 for $L_z=4h_1$ and $5h_1$ 
at $h_2/h_1=1$, respectively, and $y/h_1=0.65$ and 0.6 for $L_z=4h_1$ and $5h_1$ at $h_2/h_1=7/8$, respectively. 
These positions are the cross-sections where $u_\mathrm{rms}$ shows 
the highest turbulence value at the leg of the hairpin vortex for each $h_2/h_1$ and $L_z$. 
The central cross-section of the hill at $z/h_1<0$ is $z_c/h_1=0$. 
For all distributions, the distribution obtained from the data 
for a single hairpin vortex \citep{Yanaoka_et_al_2007b} is symmetric 
to the central cross-section of the hill. 
At $x/h_1=4$, $u_\mathrm{rms}$ for each condition is at the same level 
as that for a single hairpin vortex. 
At $x/h_1=8$, $u_\mathrm{rms}$ on the spanwise center side at $z_c/h_1> 0$ increases, 
and the distribution becomes asymmetric, 
indicating that interference between hairpin vortices occurs. 
On the spanwise center side, $-\overline{u'\theta'}$ is also high, 
indicating active heat transport. 
At that time, the values of all turbulences are higher than 
that for a single hairpin vortex.

Compared with the result of $h_2/h_1=7/8$, 
$u_\mathrm{rms}$ on the spanwise center side at $h_2/h_1=1$ is higher, 
and the distribution asymmetry is also larger. 
It can be seen from this that the interference between hairpin vortices 
with the same scale strongly appears. 
In the distribution of $-\overline{u'\theta'}$, 
the same tendency as $u_\mathrm{rms}$ can be observed, 
and heat transport on the spanwise center side at $h_2/h_1=1$ is active.

When the hill spacing is narrowed, 
the interference between hairpin vortices is more intensive, 
and $u_\mathrm{rms}$ on the spanwise center side increases. 
For $h_2/h_1=7/8$ and $L_z=4h_1$, 
as a longitudinal vortex is generated on the spanwise center side, 
the interference between the hairpin vortex and longitudinal vortex 
also increases $u_\mathrm{rms}$. 
As the value on the spanwise center side of $-\overline{u'\theta'}$ also increases, 
it can be seen that the heat transport improves 
when the interference between hairpin vortices becomes stronger.

\begin{figure}[!t]
\centering
\begin{minipage}{0.48\linewidth}
\centering
\includegraphics[trim=1.5mm 6mm 17mm 3mm, clip, width=80mm]{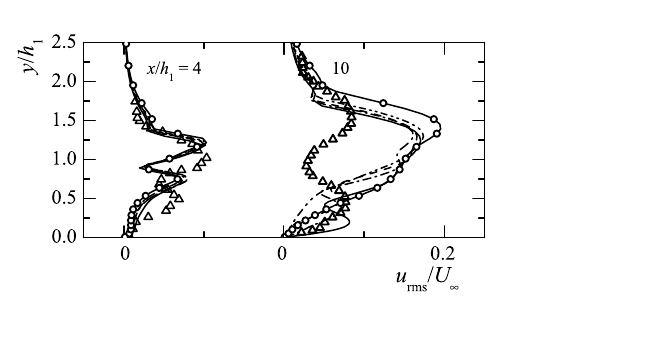} \\
\vspace*{0.5\baselineskip}
(a) $u_{\mathrm {rms}}/U_{\infty}$
\end{minipage}
\begin{minipage}{0.48\linewidth}
\vspace*{-1.0\baselineskip}
\centering
\includegraphics[trim=30mm 5mm 4mm -0.5mm, clip, width=55mm]{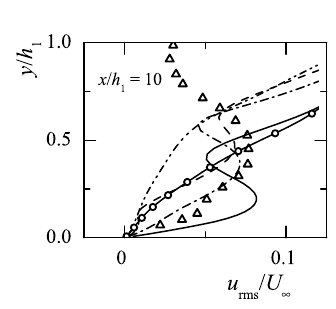} \\
(b) Enlarged view of $u_{\mathrm {rms}}/U_{\infty}$
\end{minipage}
\caption{Turbulence intensity distributions of streamwise velocity fluctuation: 
---, $h_2/h_1=1$, $L_z=4h_1$;  
\mbox{- - -,} $h_2/h_1=1$, $L_z=5h_1$; 
-- $\cdot$ --, $h_2/h_1=7/8$, $L_z=4h_1$; 
-- $\cdot$ $\cdot$ --, $h_2/h_1=7/8$, $L_z=5h_1$; 
--$\circ$--, Yanaoka et al. ($Re=500$);
$\bigtriangleup$, Acarlar--Smith ($Re=750$).}
\label{urms_xy}
\end{figure}

Figure \ref{urms_xy}(a) shows the turbulence distribution 
in the central cross-section of the hill at $z/h_1<0$. 
Figure \ref{urms_xy}(b) shows the enlarged distribution 
near the wall surface at $x/h_1=10$ in Fig. \ref{urms_xy}(a). 
The results of this calculation are compared with 
the experimental values of \citet{Acarlar&Smith_1987a} 
and the calculation results for a single hairpin vortex by the authors \citep{Yanaoka_et_al_2007b}. 
At $x/h_1=4$, two maxima are observed in this calculation for all conditions 
and in the previous results \citep{Acarlar&Smith_1987a,Yanaoka_et_al_2007b}. 
The maximum value on the mainstream side is due to the head of the hairpin vortex. 
The maximum value near the wall surface seen in the previous result \citep{Acarlar&Smith_1987a} 
is due to the leg of the hairpin vortex. 
On the other hand, as the maximum value in this calculation occurs 
near $y/h_1=0.75$ below the head of the hairpin vortex, 
this turbulence at the wall side is due to the Q2 ejection of the hairpin vortex.

At $x/h_1=10$ for $h_2/h_1=1$, 
the turbulence caused by the head of the hairpin vortex decays faster than 
the result for a single hairpin vortex, 
and a high maximum value of turbulence occurs near the wall surface. 
Then, $u_\mathrm{rms}$ for $L_z=4h_1$ has a higher value than for $L_z=5h_1$. 
When the hill spacing is narrowed, the asymmetry of the vortex structure increases 
as the interference between hairpin vortices increases. 
As a result, $u_\mathrm{rms}$ increases 
because the horn-shaped secondary vortex and the leg of the hairpin vortex 
are closer to the central cross-section of the hill.

For $h_2/h_1=7/8$ and $L_z=4h_1$, 
high turbulence appears near the wall surface at $x/h_1=10$, similar to $h_2/h_1=1$. 
This is because the horn-shaped secondary vortex and the leg of the hairpin vortex 
approach the central cross-section of the hill 
as the hairpin vortex structure becomes asymmetric 
due to the interference between hairpin vortices 
and interference between the hairpin vortex and longitudinal vortex. 
For $h_2/h_1=7/8$ and $L_z=5h_1$, 
as the influence of interference between hairpin vortices is weak, 
the asymmetry of the hairpin vortex structure is small. 
Therefore, the result of this calculation is qualitatively similar to 
that for a single hairpin vortex.

Focusing on Fig. \ref{urms_xy}(b), 
the maximum value for $h_2/h_1=1$ and $L_z=4h_1$ occurs closer to the wall surface 
than other results. 
It can be seen from this that the leg of the hairpin vortex approaches 
the wall surface as a result of the increase in the asymmetry of the hairpin vortex 
due to the interference between hairpin vortices.

\begin{figure}[!t]
\centering
\begin{minipage}{0.48\linewidth}
\centering
\includegraphics[trim=3mm 6.5mm 20mm 3.5mm, clip, width=75mm]{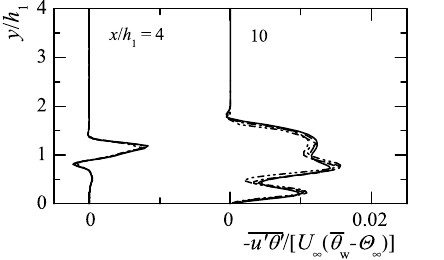} \\
(a) $h_2/h_1=1$, $L_z=4h_1$ \\
\end{minipage}
\vspace*{0.5\baselineskip}
\begin{minipage}{0.48\linewidth}
\centering
\includegraphics[trim=3mm 6.5mm 20mm 3.5mm, clip, width=75mm]{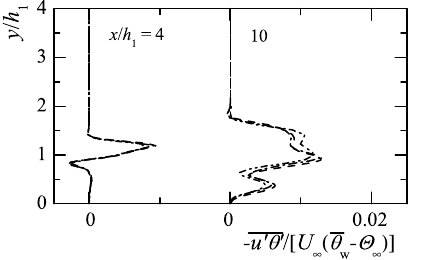} \\
(b) $h_2/h_1=7/8$, $L_z=4h_1$ \\
\end{minipage}
\caption{Distributions of turbulent heat flux: 
-- $\cdot$ $\cdot$ --, grid1; -- $\cdot$ --, grid2; - - -, grid3; ---, grid4.}
\label{ut_hikaku}
\end{figure}

Figure \ref{ut_hikaku} shows the distribution of the turbulent heat flux $-\overline{u'\theta'}$ 
obtained by the four grids for $h_2/h_1=1$ and 7/8 at $L_z=4h_1$. 
These distributions are the values in the central cross-section of the hill 
at $z/h_1<0$. 
For $h_2/h_1=1$ and $L_z=4h_1$, 
there is a difference between the result of grid4 and the results of grid1 and grid2, 
but the results of grid4 and grid3 agree well. 
For $h_2/h_1=7/8$ and $L_z=4h_1$, there is a difference between the results of grid3 and grid1, 
but the results of grid3 and grid2 agree well. 
From the above results, it is considered that grid4 and grid3 used 
in the calculations of $h_2/h_1=1$ and $7/8$ for $L_z=4h_1$ have sufficient grid resolution, respectively.

\section{Conclusions}

We performed a numerical analysis of the hairpin vortex and heat transport 
generated by the interference of the wakes behind two hills in a laminar boundary layer. 
As a result, the following results were obtained.

When hills with the same scale are installed, 
hairpin vortices with the same scale are periodically shed behind the hills 
symmetrically with respect to the spanwise center. 
The hairpin vortex becomes asymmetric downstream. 
Hairpin vortices with different scales periodically occur 
behind the different-scale hills. 
In the case of hills with the same scale, 
the interference between hairpin vortices in the wake is stronger than 
in the different-scale hills. 
When the hill spacing in the spanwise direction is narrowed, 
the asymmetry of the hairpin vortex in the wake increases 
due to the interference between the wakes.

High turbulence areas occur around the head and both legs of the hairpin vortex, 
and the turbulence due to the horn-shaped secondary vortex occurs near the wall surface. 
Regardless of the scale ratio of the hills, 
when the hill spacing is narrowed, the significant interference between hairpin vortices 
increases the turbulence due to the leg and horn-shaped secondary vortex 
on the spanwise center side in the hairpin vortex, 
resulting in active heat transport around the hairpin vortex. 
In addition, the strong interference between hairpin vortices increases 
the asymmetry of the hairpin vortex, 
and the leg of the hairpin vortex approaches the wall surface. 
As a result, the leg removes high-temperature fluid near the wall surface 
over a wide area, improving heat transfer. 
These tendencies are most remarkable in the same-scale hills.

Around the hills with different scales, if the hill spacing is narrowed, 
the vortex size on the spanwise center side in the vortex pair generated 
behind the small hill becomes smaller due to the interference between the wakes. 
As a result, as the rollup of the vortex on the spanwise center side is weakened, 
the leg on the spanwise center side in a small hairpin vortex does not develop downstream 
and decays quickly. 
Due to the attenuation of the leg, 
heat transport around the hairpin vortex decays, 
and the heat transfer coefficient significantly decreases.


\vspace*{1.0\baselineskip}
\noindent
{\bf Acknowledgements.}
The numerical results in this research were obtained 
using supercomputing resources at the Cyberscience Center, Tohoku University. 
This research did not receive any specific grant from funding agencies 
in the public, commercial, or not-for-profit sectors. 
We would like to express our gratitude to Associate Professor Yosuke Suenaga 
of Iwate University for his support of our laboratory. 
The authors wish to acknowledge the time and effort of everyone involved in this study.

\vspace*{1.0\baselineskip}
\noindent
{\bf Declaration of interests.}
The authors report no conflicts of interest.

\vspace*{1.0\baselineskip}
\noindent
{\bf Author ORCID.} \\
H. Yanaoka \url{https://orcid.org/0000-0002-4875-8174}.

\vspace*{1.0\baselineskip}
\noindent
{\bf Author contributions.}
H. Yanaoka conceived and planned the research 
and developed the computational method and numerical codes. 
T. Hamada performed the simulations. 
All authors contributed equally to analyzing data, reaching conclusions, 
and writing the paper.


\bibliographystyle{arXiv_elsarticle-harv}
\bibliography{reference_hamada_turbulence_bibfile}

\begin{thebibliography}{17}
\expandafter\ifx\csname natexlab\endcsname\relax\def\natexlab#1{#1}\fi
\providecommand{\url}[1]{\texttt{#1}}
\providecommand{\href}[2]{#2}
\providecommand{\path}[1]{#1}
\providecommand{\DOIprefix}{doi:}
\providecommand{\ArXivprefix}{arXiv:}
\providecommand{\URLprefix}{URL: }
\providecommand{\Pubmedprefix}{pmid:}
\providecommand{\doi}[1]{\href{http://dx.doi.org/#1}{\path{#1}}}
\providecommand{\Pubmed}[1]{\href{pmid:#1}{\path{#1}}}
\providecommand{\bibinfo}[2]{#2}
\ifx\xfnm\relax \def\xfnm[#1]{\unskip,\space#1}\fi
\bibitem[{Acarlar and Smith(1987)}]{Acarlar&Smith_1987a}
\bibinfo{author}{Acarlar, M.S.}, \bibinfo{author}{Smith, C.R.},
  \bibinfo{year}{1987}.
\newblock \bibinfo{title}{A study of hairpin vortices in a laminar boundary
  layer. {P}art 1. {H}airpin vortices generated by a hemisphere protuberance}.
\newblock \bibinfo{journal}{J. Fluid Mech.} \bibinfo{volume}{175},
  \bibinfo{pages}{1--41}.
\newblock \DOIprefix\doi{https://doi.org/10.1017/S0022112087000272}.
\bibitem[{Amsden and Harlow(1970)}]{Amsden&Harlow_1970}
\bibinfo{author}{Amsden, A.A.}, \bibinfo{author}{Harlow, F.H.},
  \bibinfo{year}{1970}.
\newblock \bibinfo{title}{A simplified {MAC} technique for incompressible fluid
  flow calculations}.
\newblock \bibinfo{journal}{J. Comput. Phys.} \bibinfo{volume}{6},
  \bibinfo{pages}{322--325}.
\newblock \DOIprefix\doi{https://doi.org/10.1016/0021-9991(70)90029-X}.
\bibitem[{Dong and Meng(2004)}]{Dong&Meng_2004}
\bibinfo{author}{Dong, S.}, \bibinfo{author}{Meng, H.}, \bibinfo{year}{2004}.
\newblock \bibinfo{title}{Flow past a trapezoidal tab}.
\newblock \bibinfo{journal}{J. Fluid Mech.} \bibinfo{volume}{510},
  \bibinfo{pages}{219--242}.
\newblock \DOIprefix\doi{https://doi.org/10.1017/S0022112004009486}.
\bibitem[{Elavarasan and Meng(2000)}]{Elavarasan&Meng_2000}
\bibinfo{author}{Elavarasan, R.}, \bibinfo{author}{Meng, H.},
  \bibinfo{year}{2000}.
\newblock \bibinfo{title}{Flow visualization study of role of coherent
  structures in a tab wake}.
\newblock \bibinfo{journal}{Fluid Dyn. Res.} \bibinfo{volume}{27},
  \bibinfo{pages}{183--197}.
\newblock \DOIprefix\doi{https://doi.org/10.1016/s0169-5983(00)00003-4}.
\bibitem[{Ko et~al.(1996)Ko, Wong and Leung}]{Ko_et_al_1996}
\bibinfo{author}{Ko, N.W.M.}, \bibinfo{author}{Wong, P.T.Y.},
  \bibinfo{author}{Leung, R.C.K.}, \bibinfo{year}{1996}.
\newblock \bibinfo{title}{Interaction of flow structures within bistable flow
  behind two circular cylinders of different diameters}.
\newblock \bibinfo{journal}{Exp. Therm. Fluid Sci.} \bibinfo{volume}{12},
  \bibinfo{pages}{33--44}.
\newblock \DOIprefix\doi{https://doi.org/10.1016/0894-1777(95)00065-8}.
\bibitem[{Kurita and Yahagi(2008)}]{Kurita&Yahagi_2008}
\bibinfo{author}{Kurita, E.}, \bibinfo{author}{Yahagi, Y.},
  \bibinfo{year}{2008}.
\newblock \bibinfo{title}{Vortex structure behind highly heated two cylinders
  in parallel arrangements}.
\newblock \bibinfo{journal}{JSME, Ser. B} \bibinfo{volume}{74},
  \bibinfo{pages}{1600--1608}.
\newblock \DOIprefix\doi{https://doi.org/10.1002/htj.20244}. \bibinfo{note}{(in
  Japanese)}.
\bibitem[{Li et~al.(2019)Li, Yu, Wang and Xu}]{Li_et_al_2019}
\bibinfo{author}{Li, H.}, \bibinfo{author}{Yu, T.}, \bibinfo{author}{Wang, D.},
  \bibinfo{author}{Xu, H.}, \bibinfo{year}{2019}.
\newblock \bibinfo{title}{Heat--transfer enhancing mechanisms induced by the
  coherent structures of wall--bounded turbulence in channel with rib}.
\newblock \bibinfo{journal}{Int. J. Heat Mass Transf.} \bibinfo{volume}{137},
  \bibinfo{pages}{446--460}.
\newblock
  \DOIprefix\doi{https://doi.org/10.1016/j.ijheatmasstransfer.2019.03.122}.
\bibitem[{Lighthill(1950)}]{Lighthill_1950}
\bibinfo{author}{Lighthill, M.J.}, \bibinfo{year}{1950}.
\newblock \bibinfo{title}{Contributions to the theory of heat transfer through
  a laminar boundary layer}.
\newblock \bibinfo{journal}{Proc. R. Soc. Lond. A} \bibinfo{volume}{202},
  \bibinfo{pages}{359--377}.
\newblock \DOIprefix\doi{https://doi.org/10.1098/rspa.1950.0106}.
\bibitem[{Meinders and
  Hanjali$\acute{\mbox{c}}$(2002)}]{Meinders&Hanjalic_2002}
\bibinfo{author}{Meinders, E.R.}, \bibinfo{author}{Hanjali$\acute{\mbox{c}}$,
  K.}, \bibinfo{year}{2002}.
\newblock \bibinfo{title}{Experimental study of the convective heat transfer
  from in--line and staggered configurations of two wall--mounted cubes}.
\newblock \bibinfo{journal}{Int. J. Heat Mass Transf.} \bibinfo{volume}{45},
  \bibinfo{pages}{465--482}.
\newblock \DOIprefix\doi{https://doi.org/10.1016/S0017-9310(01)00180-6}.
\bibitem[{Rhie and Chow(1983)}]{Rhie&Chow_1983}
\bibinfo{author}{Rhie, C.M.}, \bibinfo{author}{Chow, W.L.},
  \bibinfo{year}{1983}.
\newblock \bibinfo{title}{Numerical study of the turbulent flow past an airfoil
  with trailing edge separation}.
\newblock \bibinfo{journal}{AIAA J.} \bibinfo{volume}{21},
  \bibinfo{pages}{1525--1532}.
\newblock \DOIprefix\doi{https://doi.org/10.2514/3.8284}.
\bibitem[{Sakamoto et~al.(1987)Sakamoto, Haniu and Obata}]{Sakamoto_et_al_1987}
\bibinfo{author}{Sakamoto, H.}, \bibinfo{author}{Haniu, H.},
  \bibinfo{author}{Obata, Y.}, \bibinfo{year}{1987}.
\newblock \bibinfo{title}{Vortex shedding from a circular cylinder placed
  vertically in a laminar boundary layer}.
\newblock \bibinfo{journal}{JSME, Ser. B} \bibinfo{volume}{53},
  \bibinfo{pages}{714--721}.
\newblock \DOIprefix\doi{https://doi.org/10.1299/KIKAIB.53.714}.
  \bibinfo{note}{(in Japanese)}.
\bibitem[{Simpson et~al.(2002)Simpson, Long and Byun}]{Simpson_et_al_2002}
\bibinfo{author}{Simpson, R.L.}, \bibinfo{author}{Long, C.H.},
  \bibinfo{author}{Byun, G.}, \bibinfo{year}{2002}.
\newblock \bibinfo{title}{Study of vortical separation from an axisymmetric
  hill}.
\newblock \bibinfo{journal}{Int. J. Heat Fluid Flow} \bibinfo{volume}{23},
  \bibinfo{pages}{582--591}.
\newblock \DOIprefix\doi{https://doi.org/10.1016/S0142-727X(02)00154-6}.
\bibitem[{Sumner et~al.(1999)Sumner, Wong, Price and
  Paidoussis}]{Sumner_et_al_1999}
\bibinfo{author}{Sumner, D.}, \bibinfo{author}{Wong, S.S.T.},
  \bibinfo{author}{Price, S.J.}, \bibinfo{author}{Paidoussis, M.P.},
  \bibinfo{year}{1999}.
\newblock \bibinfo{title}{Fluids behavior of side--by--side circular cylinders
  in steady cross--flow}.
\newblock \bibinfo{journal}{J. Fluids Struct.} \bibinfo{volume}{13},
  \bibinfo{pages}{309--338}.
\newblock \DOIprefix\doi{https://doi.org/10.1006/jfls.1999.0205}.
\bibitem[{Yanaoka et~al.(2007a)Yanaoka, Inamura and
  Kawabe}]{Yanaoka_et_al_2007a}
\bibinfo{author}{Yanaoka, H.}, \bibinfo{author}{Inamura, T.},
  \bibinfo{author}{Kawabe, S.}, \bibinfo{year}{2007}a.
\newblock \bibinfo{title}{Numerical simulation of behavior and heat transfer of
  hairpin vortices generated around a cube in a laminal boundary layer}.
\newblock \bibinfo{journal}{JSME, Ser. B} \bibinfo{volume}{73},
  \bibinfo{pages}{268--275}.
\newblock \DOIprefix\doi{http://dx.doi.org/10.1299/kikaib.73.268}.
  \bibinfo{note}{(in Japanese)}.
\bibitem[{Yanaoka et~al.(2008)Yanaoka, Inamura, Suenaga and
  Abe}]{Yanaoka_et_al_2009}
\bibinfo{author}{Yanaoka, H.}, \bibinfo{author}{Inamura, T.},
  \bibinfo{author}{Suenaga, Y.}, \bibinfo{author}{Abe, K.},
  \bibinfo{year}{2008}.
\newblock \bibinfo{title}{Numerical analysis of the effect of boundary layer
  thickness on vortex structures and heat transfer in the wake behind a hill}.
\newblock \bibinfo{journal}{Heat Trans. Asian Res.} \bibinfo{volume}{38},
  \bibinfo{pages}{435--449}.
\newblock \DOIprefix\doi{https://doi.org/10.1002/htj.20261}.
\bibitem[{Yanaoka et~al.(2007b)Yanaoka, Inamura, Suenaga and
  Kobayashi}]{Yanaoka_et_al_2007b}
\bibinfo{author}{Yanaoka, H.}, \bibinfo{author}{Inamura, T.},
  \bibinfo{author}{Suenaga, Y.}, \bibinfo{author}{Kobayashi, Y.},
  \bibinfo{year}{2007}b.
\newblock \bibinfo{title}{Numerical simulation of vortex structures and heat
  transfer behind a hill in a laminar boundary layer}.
\newblock \bibinfo{journal}{JSME, Ser. B} \bibinfo{volume}{73},
  \bibinfo{pages}{2357--2544}.
\newblock \DOIprefix\doi{https://doi.org/10.1299/kikaib.73.2537}.
  \bibinfo{note}{(in Japanese)}.
\bibitem[{Yang et~al.(2001)Yang, Meng and Sheng}]{Yang_et_al_2001}
\bibinfo{author}{Yang, W.}, \bibinfo{author}{Meng, H.}, \bibinfo{author}{Sheng,
  J.}, \bibinfo{year}{2001}.
\newblock \bibinfo{title}{Dynamics of hairpin vortices generated by a mixing
  tab in a channel flow}.
\newblock \bibinfo{journal}{Exp. Fluids} \bibinfo{volume}{30},
  \bibinfo{pages}{705--722}.
\newblock \DOIprefix\doi{https://doi.org/10.1007/s003480000252}.

\end{thebibliography}

\end{document}